\documentclass[preprint,preprintnumbers,showpacs,aps,prd,amssymb]{revtex4-1}

\usepackage{graphicx}
\usepackage{bm}
\usepackage{amsmath}
\usepackage{xcolor}

\def\calH{{\cal H}}
\def\calL{{\cal L}}

\def\calO{{\cal O}}

\def\Bbar{{\bar B}}
\def\bbar{{\bar b}}
\def\cbar{{\bar c}}
\def\dbar{{\bar d}}

\def\hbar{{\bar h}}
\def\ellbar{{\bar\ell}}

\def\Qbar{{\bar Q}}
\def\sbar{{\bar s}}
\def\ubar{{\bar u}}
\def\nubar{{\bar\nu}}

\def\mhat{{\hat m}}
\def\shat{{\hat s}}
\def\uhat{{\hat u}}

\def\tR{{\tilde R}}

\def\avg{{\rm avg}}
\def\best{{\rm best}}
\def\Br{{\rm Br}}
\def\dof{{\rm d.o.f.}}
\def\eff{{\rm eff}}
\def\em{{\rm em}}
\def\exp{{\rm exp}}
\def\GeV{{\rm GeV}}

\def\LQ{{\rm LQ}}
\def\min{{\rm min}}
\def\NP{{\rm NP}}
\def\Re{{\rm Re}}
\def\SM{{\rm SM}}
\def\SU{{\rm SU}}
\def\TeV{{\rm TeV}}

\def\nn{\nonumber}

\def\Ds{{D^{(*)}}}
\def\Ks{{K^{(*)}}}

\def\Bs2mumu{{B_s\to\mu^+\mu^-}}
\def\B2Knunu{{B^+\to K^+\nu\nubar}}

\begin{document}
\title{$B$ anomalies and the tauphilic leptoquark model}
\author{Jong-Phil Lee}
\email{jongphil7@gmail.com}
\affiliation{Sang-Huh College,
Konkuk University, Seoul 05029, Korea}

\begin{abstract}
We suggest three scalar leptoquarks, $S_1$, ${\tilde R}_2$, and $S_3$ 
which couple exclusively to the 3rd generation of leptons to explain the $B$ anomalies.
The scalar singlet $S_1$ can explain $R(D^{(*)})$, while the doublet ${\tilde R}_2$ is needed to fit 
the branching ratio ${\rm Br}(B\to K^{(*)}\nu{\bar\nu})$ via the right-handed Wilson coefficients $C_R^\nu$.
A strong constraint from $\Delta m_{B_s}$ requires triplet $S_3$ where the mixing term between
$S_1$ and $S_3$ is important.
Our analysis suggests subdominant $C_{V_L}<0$ and dominant $C_{S_L} >0$ and 
lepton-flavor universality violating $C_9^{\rm LQ} \approx +1$.
The masses are expected to be $M_{S_1}\sim M_{{\tilde R}_2} < M_{S_3} \lesssim 3$ TeV,
which can be probed in future colliders.
\end{abstract}

\maketitle
\section{Introduction}
%
\par
$B$ anomalies in recent years provide challenges for the validity of the standard model (SM) and probes for 
new physics (NP).
The anomalies include $R(\Ds)$ in $b\to c$ transition,
\begin{equation}
R(\Ds)\equiv\frac{{\rm Br}(B\to \Ds\tau\nu)}{{\rm Br}(B\to \Ds\ell\nu)}~,
\label{RDsDef}
\end{equation}
where $\ell =\mu$ or $e$.
The SM predictions for these observables are \cite{DAlise2403,Iguro2405}
\begin{eqnarray}
R(D)_\SM &=& 0.290\pm0.003~, \label{RDSM} \\
R(D^*)_\SM &=& 0.248\pm0.001~. \label{RDsSM} 
\end{eqnarray}
\par
Experimentally, there are some tensions compared with the SM predictions.
See Table \ref{T_RDs} for experimental data.
We have the averaged values from the measurements 
\cite{BaBar1,Belle1,Belle1607,Belle1703,Belle1612, Belle1709,Belle1904,BelleII2024,
LHCb2302,LHCb2305,LHCb24} 
as
\begin{eqnarray}
R(D)_\avg &=& 0.339\pm0.026~, ~~~\\
R(D^*)_\avg &=& 0.285\pm0.011~.
\end{eqnarray}
The Pull values with respect to the SM are $1.87$ for $R(D)$ and $3.35$ for $R(D^*)$,
suggesting the existence of some NP scenarios for the lepton flavor universality (LFU) violation.
%
\begin{table}
\begin{tabular}{|c| ccc |}\hline\hline
              & ~$R(D)$ & ~$R(D^*)$   & correlation \\\hline
 BABAR \cite{BaBar1} & ~$0.440\pm0.058\pm0.042~~~$      & ~$0.332\pm0.024\pm0.018$ & $-0.27$  \\
 Belle(2015) \cite{Belle1}  & ~$0.375\pm0.064\pm0.026~~~$ & ~$0.293\pm0.038\pm0.015$ & $-0.49$ \\
 Belle(2016) \cite{Belle1607} & ~$-$                                        & ~$0.302\pm0.030\pm0.011$ &  $-$ \\
 Belle(2017) \cite{Belle1703} & ~$-$                                        & ~$0.276\pm0.034^{+0.029}_{-0.026}$ & $-$  \\
 Belle(2017) \cite{Belle1612,Belle1709} & ~$-$                       & ~$0.270\pm0.035^{+0.028}_{-0.025}$ & $-$ \\
 Belle(2019) \cite{Belle1904} & ~$0.307\pm0.037\pm0.016~~~$ & ~$0.283\pm0.018\pm0.014$ & $-0.51$ \\
 BelleII(2024)  \cite{BelleII2024} & ~$-$                                        & ~$0.262^{+0.041+0.035}_{-0.039-0.032}$ & $-$ \\
 LHCb(2023) \cite{LHCb2302} & $0.441\pm 0.060 \pm 0.066~~~$ & $0.281\pm0.018\pm0.024$ & $-0.43$ \\
 LHCb(2023) \cite{LHCb2305} & ~$-$                                           & ~$0.267\pm0.012\pm0.019$ & $-$ \\
 LHCb(2024) \cite{LHCb24} & ~$0.249\pm0.043\pm0.047~~~$   & ~$0.402\pm0.081\pm0.085$  & $-0.39$\\ 
 \hline\hline
 \end{tabular}
\caption{Summary of experimental data for $R(\Ds)$.
The uncertainties are $\pm$(statistical)$\pm$(systematic).
The correlations between $R(D)$ and $R(D^*)$  are also provided \cite{LHCb24,HFAG2019}.}
\label{T_RDs}
\end{table}
%
%
%
\par
The decay mode of $B\to\Ds\tau\nubar$ also contains information about polarizations.
For the decay width $\Gamma_\tau^\Ds(\pm)$ where $(\pm)$ corresponds to $(\pm)~\tau$ helicity,
one can define the polarization asymmetry of $\tau$ as
\begin{equation}
P_\tau(\Ds) \equiv \frac{\Gamma^{D^{(*)}}_\tau(+)-\Gamma^{D^{(*)}}_\tau(-)}
                                          {\Gamma^{D^{(*)}}_\tau(+)+\Gamma^{D^{(*)}}_\tau(-)}~.
\label{PDsDef}
\end{equation}
Regarding $D^*$ a fraction of the $D^*$-longitudinal mode defined by
\begin{equation}
F_L(D^*)\equiv\frac{\Gamma(B\to D_L^*\tau\nu)}{\Gamma(B\to D^*\tau\nu)}~,
\label{FLDef}
\end{equation}
is a useful quantity.
The SM estimations are \cite{DAlise2403,Iguro2405}
\begin{eqnarray}
P_\tau(D)_\SM &=& 0.331\pm0.004~, \label{PDSM} \\
P_\tau(D^*)_\SM &=& -0.497\pm0.007~,\label{PDsSM} \\
F_L(D^*)_\SM &=& 0.464\pm0.003~. \label{FLSM}
\end{eqnarray}
\par
The $\tau$-polarization asymmetry was measured to be \cite{Belle1612,Belle1709}
\begin{equation}
P_\tau(D^*) = -0.38\pm 0.51^{+0.21}_{-0.16}~,
\end{equation}
where the uncertainties are somewhat large.
Previous measurement of the longitudinal $D^*$ polarization by the Belle collaboration\cite{Belle1903}, 
$F_L(D^*) = 0.60\pm0.08\pm0.04$ was rather larger than the SM prediction.
However, recent measurement the LHCb \cite{LHCb2311},
$F_L(D^*) = 0.43\pm0.06\pm0.03$ is much smaller.
The combined result of
\begin{equation}
F_L(D^*) = 0.49\pm0.05~,
\end{equation}
is consistent with SM value of Eq.\ (\ref{FLSM}).
Related experimental data are summarized in Table \ref{T_polDs}.
The $\tau$-polarization asymmetry for $B\to D\tau\nubar$ decay, $P_\tau(D)$, has not been measured so far.
%
\begin{table}
\begin{tabular}{|c| cc |}\hline\hline
 & $P_\tau(D^*)$ & $F_L(D^*)$  \\\hline
 Belle(2017) & $-0.38\pm0.51^{+0.21}_{-0.16}$\cite{Belle1612,Belle1709} & $-$  \\
 Belle(2019)  & ~$-$ & ~$0.60\pm0.08\pm0.04$ \cite{Belle1903}   \\
LHCb(2023) & ~$-$ & ~$0.43\pm0.06\pm0.03$ \cite{LHCb2311}   \\
 \hline\hline
 \end{tabular}
\caption{Summary of experimental data for $P_\tau(\Ds)$ and $F_L(D^*)$.
}
\label{T_polDs}
\end{table}
%
%
%
%
\par
Another observables of $b\to c\ell\nubar$ decay mode involve $B_c$ meson and $\Lambda_c$ baryon.
Similar to $R(\Ds)$, one can define $R(J/\Psi)$ and $R(\Lambda_c)$ as
%
%
%
%
%
%
\begin{eqnarray}
R(J/\Psi) &=& \frac{\Br(B_c\to J/\Psi\tau\nubar)}{\Br(B_c\to J/\Psi\mu\nubar)}~,\label{RJPDef}\\
R(\Lambda_c) &=& \frac{\Br(\Lambda_b\to\Lambda_c\tau\nubar)}
                                       {\Br(\Lambda_b\to\Lambda_c\ell\nubar)}~.\label{RLdcDef}
\end{eqnarray}
The SM predictions are \cite{LQCD2007,Bernlochner1808,Bernlochner1812} 
\begin{eqnarray}
R(J/\Psi)_\SM &=& 0.258\pm0.004 ~,\nn\\
R(\Lambda_c)_\SM& =& 0.324\pm0.004 ~.
\end{eqnarray}
Experimental data for $R(J/\Psi)$ and $R(\Lambda_c)$ are given in Table \ref{T_RJRLc}.
The averaged value of $R(J/\Psi)$ is
$R(J/\Psi) = 0.61\pm 0.18 $ \cite{Iguro2405}, 
and both $R(J/\Psi)$ and $R(\Lambda_c)$ reveal some tension with the SM results. 
NP effects, if exist, seem to enhance $R(J/\Psi)$ while reduce $R(\Lambda_c)$.
Though the experimental uncertainties are large in $R(J/\Psi)$, it could be complementary to $R(\Ds)$.
Also, interestingly enough, $R(\Lambda_c)$ is known to satisfy a sum rule with respect to $R(\Ds)$ as
\cite{Blanke1811,Blanke1905,Fedele2211}
\begin{equation}
\frac{R(\Lambda_c)}{R(\Lambda_c)_\SM} \simeq 
0.280\frac{R(D)}{R(D)_\SM} + 0.720\frac{R(D^*)}{R(D^*)_\SM}~.
\label{sumrule}
\end{equation}
The sum rule connects the baryonic and mesonic modes and would be helpful to check 
nontrivial NP effects on different hadronic structure.
%
\begin{table}
\begin{tabular}{cc}\hline\hline
$R(J/\Psi)$                                                                      &  $R(\Lambda_c)$ \\ \hline
$~0.71\pm0.17\pm0.18~~~$ \cite{LHCb1711} & $0.242\pm0.026\pm0.071$  \cite{LHCb2201} \\
$~0.17\pm0.33~~~$ \cite{CMS2408} & \\
$~1.04^{+0.50}_{-0.44}~~~$ \cite{CMS2510} & \\ \hline\hline
\end{tabular}
\caption{\label{T_RJRLc} Experimental data for $R(J/\Psi)$ and $R(\Lambda_c)$.}
\end{table}
%
\par
Another important observable of $b\to c\tau\nubar$ is the branching ratio of $\Br(B_c^-\to\tau^-\nubar)$.
Though the branching ratio has not be observed, it puts constraints on NP scenarios.
The SM prediction is $\Br(B_c^-\to\tau^-\nubar)_\SM = (2.29\pm0.09)\%$ \cite{Zuo2305}.
Usually the upper bound  of the branching ratio is imposed to be several tens of percent.
In this analysis we adopt a moderate bound of $\Br(B_c^-\to\tau^-\nu)<0.3$ \cite{Alonso1611}.
%
%
%
%
\par
In the $b\to s$ transition modes similar fractions to $R(\Ds)$ are
\begin{equation}
R(\Ks)\equiv\frac{\Br(B\to\Ks\mu^+\mu^-)}{\Br(B\to\Ks e^+e^-)}~.
\end{equation}
Previous experimental data exhibited some tension with the SM, suggesting possible lepton-universality violations
\cite{LHCb1705,LHCb2103,Geng2103}.
But the tension diminished after the LHCb measurements \cite{LHCb2212_52,LHCb2212_53}.
The results are in good agreement with the SM calculations,
which are very close to unity \cite{Hiller0310,Bobeth0709,Geng1704,Bordone1605}.
However, other observables like $\Br(B\to \Ks\mu^+\mu^-)$, $\Br(B_s\to\phi\mu^+\mu^-)$, and
the angular observable $P_5^\prime(K^*)$ still deviate from the SM estimations.
In case of $b\to s\tau^+\tau^-$ decay modes, experimental searches provide upper bounds as \cite{PDG2024}
\begin{eqnarray}
\Br(B^+\to K^+\tau^+\tau^-) < 2.25\times 10^{-3}~,\nn\\
\Br(B^0\to K^{*0}\tau^+\tau^-) < 3.1\times 10^{-3}~.
\end{eqnarray}
The results are still far above the SM calculations \cite{Bause2109,Bause2309}
\begin{eqnarray}
\Br(B^+\to K^+\tau^+\tau^-)_\SM = (1.17\pm0.14)\times 10^{-7}~,\nn\\
\Br(B^0\to K^{*0}\tau^+\tau^-)_\SM= (0.87\pm0.11)\times 10^{-7}~.
\end{eqnarray}
Related decay mode is $B_s\to \tau^+\tau^-$.
The SM predicts \cite{Capdevila1712}
\begin{equation}
\Br(B_s\to\tau^+\tau^-)_\SM = (7.73\pm0.49)\times 10^{-7}~,
\end{equation}
and the experimental bound is \cite{PDG2024}
\begin{equation}
\Br(B_s\to\tau^+\tau^-) < 6.8 \times 10^{-3}~.
\end{equation}
The experimental upper bound is very far from the SM estimations.
As in cases of $B\to\Ks\tau^+\tau^-$, experimental upper bounds are $\calO(10^{-3})$ 
while the SM predictions are $\calO(10^{-7})$.
%
%
%
%
\par
Dineutrino decay modes of $B\to\Ks\nu\nubar$ are also very interesting.
Recently the Belle II collaboration announced a large branching ratio \cite{Belle2_2311},
\begin{equation}
\Br(\B2Knunu)_\exp = (2.3\pm 0.7)\times 10^{-5}~.
\end{equation}
The SM predicts much smaller value \cite{Becirevic2301},
\begin{equation}
\Br(\B2Knunu)_\SM = (4.43\pm0.31)\times 10^{-6}~,
\end{equation}
calling for NP explanations.
Meanwhile, $B^0\to K^{*0}\nu\nubar$ is not yet measured
and the upper bound by the Belle collaboration search is  \cite{Belle1702}
\begin{equation}
\Br(B^0\to K^{*0}\nu\nubar) < 1.8\times 10^{-5}~.
\label{ctr_Ks}
\end{equation}
The SM estimation \cite{Becirevic2301},
\begin{equation}
\Br(B^0\to K^{*0}\nu\nubar)_\SM = (9.47\pm1.40)\times 10^{-6}~,
\end{equation}
lies below the experimental bound.
%
%
%
\par
Another observable we must consider is $\Delta m_{B_s}$ in $B_s$-$\Bbar_s$ mixing.
Experimentally $\Delta m_{B_s}$ is well measured \cite{CMS2007,LHCb2011,LHCb2104,LHCb2308} and
can provide strong constraints on NP.
In Table \ref{T_b2s} measurements of $\Delta m_{B_s}$ as well as $\Br(B^+\to K^+\nu\nubar)$ are summarized.
The SM estimation \cite{Luzio1909},
\begin{equation}
\Delta m_{B_s}^\SM = (18.4^{+0.7}_{-1.2})\times 10^{12}\cdot {\rm sec}^{-1}~,
\end{equation}
is slightly higher than the experimental data.
%
%
%
\begin{table}
\begin{tabular}{cc}\hline\hline
$\Delta m_{B_s}\times 10^{-12}\cdot{\rm sec}$ &  $\Br(B^+\to K^+\nu\nubar)$ \\ \hline
$~17.51^{+0.10}_{-0.09}\pm0.03~~~$ \cite{CMS2007} & $(2.3\pm0.7)\times 10^{-5}$  \cite{Belle2_2311} \\
$~17.757\pm0.007\pm0.008~~~$ \cite{LHCb2011} & \\
$~17.7683\pm0.0051\pm0.0032~~~$ \cite{LHCb2104} & \\
$~17.743\pm0.033\pm0.009~~~$ \cite{LHCb2308} \\ \hline\hline
\end{tabular}
\caption{\label{T_b2s} Experimental data for $\Delta m_{B_s}$ and $\Br(B^+\to K^+\nu\nubar)$.}
\end{table}
%
%
%
%
\par
So far there have been many NP scenarios to explain the $B$ anomalies.
Some examples include
supersymmetry (SUSY) \cite{Altmannshofer2002,Bardhan2107,Zheng2410}, 
$Z'$ \cite{Crivellin1501,Crivellin1503,Chiang1706,King1706,Chivukula1706,Cen2104,Davighi2105}, 
new scalars \cite{Hu1612,Crivellin1903,Rose1903,Ho2401}, and
unparticles \cite{JPL2106}, etc.
In more model-independent ways, the relevant Wilson coefficients can be parameterized generically 
as $C_j^\NP\sim A_j (v/M_\NP)^\alpha$,
where $v$ is the SM vacuum expectation value, $M_\NP$ is the NP scale, and $A_j$ is the combination of
fermionic couplings \cite{JPL2110,JPL2208,JPL2411,JPL2502,JPL2509}.
Inspired by unparticle scenario, $\alpha$ can be non integers.
\par
Among the specific NP scenario is the leptoquark (LQ) model which has been one of the most promising NP
\cite{Hiller1408,Becirevic1503,Dorsner1603,Bauer1511,Chen1703,Crivellin1703,Cai1704,Calibbi1709,
Blanke1801,Nomura2104,Angelescu2103,Du2104,Marzocca2106,Cheung2204,Bhaskar2204,
Chen2205,Becirevic2404,Crivellin2505}.
In this paper we introduce three scalar leptoquarks with special focus on the tauonic sector.
Our framework includes $\SU(2)_L$ singlet LQ $S_1 ({\bar 3}, 1, 1/3)$, doublet LQ $\tR_2 (3, 2, 1/6)$, and
triplet $S_3 ({\bar 3}, 3, 1/3)$. 
In order to explain $R(\Ds)_\exp > R(\Ds)_\SM$ new Wilson coefficients other than the SM value (=1) are required.
Here $S_1$ and $S_3$ can do the work via the Wilson coefficients $C_{V_L}$, $C_{S_L}$ and $C_T$.
In $b\to s\ell^+\ell^-$ decay modes, one needs $S_3$ and/or $\tR_2$.
$S_1$ can contribute through the charm loops, but subdominant.
Doublet $\tR_2$ is also needed for $b\to s\nu\nubar$ modes.
As discussed in \cite{Chen2401}, nonzero $C_R^{\nu_\ell}\ne 0$ is necessary to explain $\Br(B\to\Ks\nu\nubar)$.
$\tR_2$ naturally introduces $C_R^{\nu_\ell}$.
Finally, $\Delta m_{B_s}$ in $B_s$-$\Bbar_s$ mixing constraints the NP parameters severely. 
To reconcile the fact that $\Delta m_{B_s}^\exp \lesssim \Delta m_{B_s}^\SM$,
destructive NP contribution is very helpful. 
The mixing term of $S_1$ and $S_3$ can do the job.
Individual LQs tend to contribute constructively for real couplings.
%
%
%
%
%
\par
In this analysis we confine ourselves to the tauphilic LQs where all the LQs couple only to 
$\tau$ and $\nu_\tau$ in the leptonic sector, as in \cite{Bause2309, Crivellin2505}.
The reason is that the $\tau$-sector is very special in $R(\Ds)$ compared to $R(\Ks)$,
and analyses combined by SMEFT in \cite{Chen2401} show that 
$C_R^{\nu_e} = C_R^{\nu_\mu} =0$ while $C_R^{\nu_\tau}\ne 0$.
It also simplifies the coupling structures.
\par
The paper is organized as follows.
In the next section we summarize the LQ model of $S_1$, $\tR_2$, and $S_3$.
Relevant Wilson coefficients for various observables are provided in terms of the model parameters.
Section III discusses our results.
In the last section we conclude.
%
%
%
%
%
\section{Leptoquark model}
%
%
%
%
%
%
We introduce three scalar LQs $S_1$, $\tR_2$ and $S_3$ with the Lagrangian 
\cite{DAlise2403,Chen1703,Chen2205,Crivellin1912}
%
%
%
%
%
%
\begin{eqnarray}
\calL_{S_1+\tR_2+S_3} &=&
     - x^R_{ij}\ubar^{C,i}_R e_R^j S_1
     - x^L_{ij}\Qbar^{C,i}_L(i\sigma^2)L_L^j S_1
     - y_{ij}\dbar_R^i\tR_2^a (i\sigma^2)^{ab} L_L^{j,b} \nn\\
&&- z_{ij}\Qbar^{C,ia}(i\sigma^2)^{ab}\left({\vec\sigma}\cdot{\vec S}\right)^{bc} L_L^{j,c} + h.c.~.
\end{eqnarray}
For tauphilic model we assume that only $x_{i3}^{L,R}$, $y_{i3}$, and $z_{i3}$ are nonzero.
\par
The effective Hamiltonian for $b\to c\ell\nubar$ processes is
\begin{eqnarray}
\calH_\eff &=&\frac{4G_F}{\sqrt{2}}V_{cb}\Big[
(1+C_{V_L})\calO_{V_L} + C_{V_R}\calO_{V_R}  \nn\\
&& C_{S_L}\calO_{S_L} + C_{S_R}\calO_{S_R} + C_T\calO_T\Big],
\end{eqnarray}
where
\begin{eqnarray}
\calO_{V_L} &=& (\cbar\gamma^\mu P_L b)(\ellbar\gamma_\mu P_L\nu)~,\nn\\
\calO_{V_R} &=& (\cbar\gamma^\mu P_R b)(\ellbar\gamma_\mu P_L\nu)~,\nn\\
\calO_{S_L} &=& (\cbar P_L b)(\ellbar P_L\nu)~,\nn\\
\calO_{V_L} &=& (\cbar P_R b)(\ellbar P_L\nu)~,\nn\\
\calO_T &=& (\cbar\sigma^{\mu\nu} P_L b)(\ellbar\sigma_{\mu\nu} P_L\nu)~,\nn\\
\end{eqnarray}
and $P_{L,R}\equiv (1\mp\gamma_5)/2$.
The Wilson coefficients are \cite{DAlise2403,Crivellin1912}
\begin{eqnarray}
C_{V_L} &=& \frac{1}{4\sqrt{2}G_F V_{cb}}\left[
\frac{(Vx^{L*})_{23} x^L_{33}}{M_{S_1}^2} - \frac{(Vz^*)_{23} z_{33}}{M_{S_3}^2}\right]~,
\label{E_CVL} \nn\\
C_{S_L} &=& -4C_T 
= -\frac{1}{4\sqrt{2} G_F V_{cb}}\frac{x^{R*}_{23} x^L_{33}}{M_{S1}^2}~, \label{E_CSL}
\end{eqnarray}
where $V$ is the CKM matrix.
The renormalization group (RG) running results in the following relations \cite{Crivellin1912}:
\begin{eqnarray}
C_{V_L}(\mu_b) &=& C_{V_L}(1\TeV)~,\nn\\
\begin{pmatrix}
C_{S_L}(\mu_b)\\ C_T(\mu_b)
\end{pmatrix}
&=&
\begin{pmatrix}
1.752 & -0.287 \\
-0.004 & 0.842
\end{pmatrix}
\begin{pmatrix}
C_{S_L}(1\TeV) \\ C_T(1\TeV)
\end{pmatrix}
\end{eqnarray}
\par
%
%
%
%
Relevant observables for $b\to c\ell\nubar$ are $R(\Ds)$, $P_\tau(\Ds)$, $F_L(D^*)$, and 
$\Br(B_c\to\tau\nubar)$.
We adopt recently updated formulae for the observables as follows (at $\mu=\mu_b=4.18~\GeV$ scale) \cite{Blanke1811,Iguro2405,Aoki,Bernlochner}
\begin{eqnarray}
\label{RDs}
\frac{R(D)}{R(D)_\SM} &=&
|1+C_{V_L}|^2 + 1.01|C_{S_L}+C_{S_R}|^2 + 0.84|C_T|^2 \nn\\
&&
+ 1.49\Re\left[(1+C_{V_L})(C_{S_L}+C_{S_R})^*\right]
+ 1.08\Re\left[(1+C_{V_L})C_T^*\right] ~,\label{RDnum}\\
\frac{R(D^*)}{R(D^*)_\SM} &=& 
|1+C_{V_L}|^2 + 0.04|C_{S_L}-C_{S_R}|^2 + 16.0|C_T|^2 \nn\\
&&
- 0.11\Re\left[(1+C_{V_L})(C_{S_L}-C_{S_R})^*\right]
- 5.17\Re\left[(1+C_{V_L})C_T^*\right] ~,\label{RDsnum}
\end{eqnarray}
and
\begin{eqnarray}
\frac{P_\tau(D)}{P_\tau(D)_\SM} &=& 
\left[\frac{R(D)}{R(D)_\SM}\right]^{-1}\Big\{
|1+C_ {V_L}|^2 + 3.04|C_{S_L}+C_{S_R}|^2 + 0.17|C_T|^2 \label{PtD} \nn\\
&& 
+ 4.50\Re\left[(1+C_{V_L})(C_{S_L}+C_{S_R})^*\right]
- 1.09\Re\left(1+C_{V_L})C_T^*\right] 
 \Big\}~,\label{PDnum}\\
\frac{P_\tau(D^*)}{P_\tau(D^*)_\SM} &=& 
\left[\frac{R(D^*)}{R(D^*)_\SM}\right]^{-1}\Big\{
|1+C_ {V_L}|^2 - 0.07|C_{S_L}-C_{S_R}|^2 - 1.85|C_T|^2\nn\\
&& 
+ 0.23\Re\left[(1+C_{V_L})(C_{S_L}-C_{S_R})^*\right] 
- 3.47\Re\left[(1+C_{V_L})C_T^*\right]
 \Big\}~,\label{PDsnum}\\
\frac{F_L(D^*)}{F_L(D^*)_\SM} &=&
\left[\frac{R(D^*)}{R(D^*)_\SM}\right]^{-1}\Big\{
|1+C_ {V_L}|^2 + 0.08|C_{S_L}-C_{S_R}|^2 + 6.90|C_T|^2\nn\\
&& 
- 0.25\Re\left[(1+C_{V_L})(C_{S_L}-C_{S_R})^*\right] 
- 4.30\Re\left[(1+C_{V_L})C_T^*\right]
 \Big\}~.
\label{FLnum}
\end{eqnarray}
%
%
%
%
\par
As for $B_c\to J/\Psi\tau\nu$ and $\Lambda_b\to\Lambda_c\tau\nu$ decays, we have \cite{Iguro2405}
\begin{eqnarray}
\frac{R(J/\Psi)}{R(J/\Psi)_\SM} &=& 
|1+C_{V_L}|^2 + 0.04|C_{S_L}-C_{S_R}|^2
-0.10\Re\left[(1+C_{V_L})(C_{S_L}-C_{S_R})^*\right] \nn\\
&&
-5.39\Re\left[(1+C_{V_L})C_T^*\right] + 14.7|C_T|^2~, \label{RJPnum} \\
\frac{R(\Lambda_c)}{R(\Lambda_c)_\SM} &=&
|1+C_{V_L}|^2 + 0.32\left(|C_{S_L}|^2+|C_{S_R}|^2\right)
+0.52\Re\left[C_{S_L}C_{S_R}^*\right] \nn\\
&&
+0.50\Re\left[(1+C_{V_L})C_{S_R}^*\right] 
+0.33\Re\left[(1+C_{V_L})C_{S_L}^*\right] \nn\\
&&
-3.11\Re\left[(1+C_{V_L})C_T^*\right] + 10.4|C_T|^2~.
\label{RLdcnum}
\end{eqnarray}
%
%
%
%
\par
The branching ratio of $B_c\to\tau\nubar$ is also directly related to $C_{ij}$ as \cite{Iguro2405,DAlise2403}
\begin{equation}
\Br(B_c\to\tau\nubar)=\Br(B_c\to\tau\nubar)_\SM\Big|
1+C_{V_L}-4.35(C_{S_L}-C_{S_R})\Big|^2~,
\end{equation}
where $\Br(B_c\to\tau\nubar)_\SM = 2.29\times 10^{-2}$.
We require a moderate bound of $\Br(B_c\to\tau\nubar)<0.3$.
%
%
%
%
%
%
%
\par
Now we go to $b\to s$ transitions.
The effective Hamiltonian for $b\to s\ell^+\ell^-$ is
\begin{equation}
\calH_\eff = -\frac{4G_F}{\sqrt{2}}V_{tb}V_{ts}^*\left(\frac{\alpha_\em}{\pi}\right)\Big[
  C_9 \calO_9 + C_{10} \calO_{10}
+C_9^\prime \calO_9^\prime + C_{10}^\prime \calO_{10}^\prime\Big]~,
\end{equation}
where
\begin{eqnarray}
\calO_9^{(\prime)} &=& (\sbar\gamma^\mu P_{L(R)} b) (\ellbar\gamma_\mu\ell)~,\nn\\
\calO_{10}^{(\prime)} &=& (\sbar\gamma^\mu P_{L(R)} b) (\ellbar\gamma_\mu\gamma_5\ell)~,
\end{eqnarray}
and $C_{9,10} = C_{9,10}^\SM + C_{9,10}^\LQ$.
The leptoquarks contribute as \cite{Becirevic1503,Crivellin1912}
\begin{eqnarray}
C_9^\LQ &=& -C_{10}^\LQ = 
\frac{1}{\sqrt{2} G_F V_{tb} V_{ts}^*}\left(\frac{\pi}{\alpha_\em}\right)
\frac{z_{23}^* z_{33}}{M_{S_3}^2}~,\nn\\
C_9^\prime &=& -C_{10}^\prime = 
-\frac{1}{2\sqrt{2} G_F V_{tb} V_{ts}^*}\left(\frac{\pi}{\alpha_\em}\right)
\frac{y_{23} y_{33}^*}{M_{\tR2}^2}~.
\end{eqnarray}
%
%
%
%
For the 3rd generation leptons, we consider the branching ratios of 
$B^+\to K^+\tau^+\tau^-$ and $B^0\to K^{0*}\tau^+\tau^-$.
The matrix elements for $B\to\Ks$ are \cite{Ali99}
\begin{eqnarray}
\langle K(p)|\sbar\gamma_\mu b|B(p_B)\rangle&=&
f_+\left[(p_B+p)_\mu-\frac{m_B^2-m_K^2}{s}q_\mu\right]+\frac{m_B^2-m_K^2}{s}f_0 q_\mu~,\\
\langle K(p)|\sbar\sigma_{\mu\nu} q^\nu(1+\gamma_5)b|B(p_B)\rangle&=&
i\left[(p_B+p)_\mu s -q_\mu(m_B^2-m_K^2)\right]\frac{f_T}{m_B+m_K}~,\\
\langle K^*(p)|(V-A)_\mu|B(p_B)\rangle&=&
-i\epsilon_\mu^*(m_B+m_{K^*})A_1+i(p_B+p)_\mu(\epsilon^*\cdot p_B)\frac{A_2}{m_B+m_{K^*}}\nn\\
&&
+iq_\mu(\epsilon^*\cdot p_B)\frac{2m_{K^*}}{s}(A_3-A_0)
+\frac{\epsilon_{\mu\nu\rho\sigma}\epsilon^{*\nu}p_B^\rho b^\sigma}{m_B+m_{K^*}}2V~,
\end{eqnarray}
where $f_{+,0,T}(s), ~A_{0,1,2}(s), ~T_{1,2,3}(s), V(s)$, and $f_-=(f_0-f_+)(1-\mhat_K^2)/\shat$ are the form factors.
As for the explicit forms of the form factors, we adopt the results of \cite{Becirevic2301}.
Here,
\begin{eqnarray}
 q&=&p_B-p~,~~~s=q^2=(p_B-p)^2~,\\
 \shat&=&\frac{s}{m_B^2}~,~~~\mhat_i=\frac{m_i}{m_B}~.
 \end{eqnarray}
%
%
\par
The differential decay rates for $B\to\Ks\ell^+\ell^-$ with respect to $s$ are given by \cite{Chang2010}
\begin{eqnarray}
\frac{d\Gamma_K}{d\shat}&=&
\frac{G_F^2\alpha^2m_B^5}{2^{10}\pi^5}|V_{tb}V_{ts}^*|^2\uhat_{K,\ell}\left\{
(|A'|^2+|C'|^2)\left(\lambda_K-\frac{\uhat_{K,\ell}^2}{3}\right)
+|C'|^24\mhat_\ell^2(2+2\mhat_K^2-\shat) \right .\nn\\
&&\left. +{\rm Re}(C'D'^*)8\mhat_\ell^2(1-\mhat_K^2)+|D'|^24\mhat_\ell^2\shat\right\}~,
\\
\frac{d\Gamma_{K^*}}{d\shat}&=&
\frac{G_F^2\alpha^2m_B^5}{2^{10}\pi^5}|V_{tb}V_{ts}^*|^2\uhat_{K^*,\ell}\left\{
\frac{|A|^2}{3}\shat\lambda_{K^*}\left(1+\frac{2\mhat_\ell^2}{\shat}\right)
+|E|^2\shat\frac{\uhat_{K^*,\ell}^2}{3}\right.\nn\\
&&
+\frac{|B|^2}{4\mhat_{K^*}^2}\left[\lambda_{K^*}-\frac{\uhat_{K^*,\ell}t^2}{3}+8\mhat^2_{K^*}(\shat+2\mhat_\ell^2)\right]
+\frac{|F|^2}{4\mhat_{K^*}^2}\left[\lambda_{K^*}-\frac{\uhat_{K^*,\ell}^2}{3}+8\mhat_{K^*}^2(\shat-4\mhat_\ell^2)\right]\nn\\
&&
+\frac{\lambda_{K^*}|C|^2}{4\mhat_{K^*}^2}\left(\lambda_{K^*}-\frac{\uhat_{K^*,\ell}^2}{3}\right)
+\frac{\lambda|_{K^*}|G|^2}{4\mhat_{K^*}^2}\left[\lambda_{K^*}-\frac{\uhat_{K^*,\ell}^2}{3}
	+4\mhat_\ell^2(2+2\mhat_{K^*}^2-\shat)\right]\nn\\
&&
-\frac{{\rm Re}(BC^*)}{2\mhat_{K^*}^2}\left(\lambda_{K^*}-\frac{\uhat_{K^*,\ell}^2}{3}\right)(1-\mhat_{K^*}^2-\shat)\nn\\
&&
-\frac{{\rm Re}(FG^*)}{2\mhat_{K^*}^2}\left[\left(\lambda_{K^*}-\frac{\uhat_{K^*,\ell}^2}{3}\right)(1-\mhat_{K^*}^2
	-\shat)-4\mhat_\ell^2\lambda_{K^*}\right]\nn\\
&&\left.
-\frac{2\mhat_\ell^2}{\mhat_{K^*}^2}\lambda_{K^*}\left[{\rm Re}(FH^*)-{\rm Re}(GH^*)(1-\mhat_{K^*}^2)\right]
+\frac{\mhat_\ell^2}{\mhat_{K^*}^2}\shat\lambda_{K^*}|H|^2\right\}~,
\end{eqnarray}
where the kinematic variables are
\begin{eqnarray}
\lambda_H&=&1+\mhat_H^4+\shat^2-2\shat-2\mhat_H^2(1+\shat)~,\\
\uhat_{H,\ell}&=&\sqrt{\lambda_H\left(1-\frac{4\mhat_\ell^2}{\shat}\right)}~.
\end{eqnarray}
Here $A',\cdots, D'$ and $A,\cdots, H$ are the auxiliary functions.
They are defined by the form factors combined with the Wilson coefficients as \cite{Ali99},
%
\begin{eqnarray}
A'&=&C_9 f_+ +\frac{2\mhat_b}{1+\mhat_K}C_7^\eff f_T ~,\\
B'&=&C_9 f_- -\frac{2\mhat_b}{\shat}(1-\mhat_K)C_7^\eff f_T ~,\\
C'&=&C_{10} f_+ ~,\\
D'&=&C_{10} f_- ~,
\end{eqnarray}
and
\begin{eqnarray}
A&=&\frac{2}{1+\mhat_{K^*}}C_9 V+\frac{4\mhat_b}{\shat}C_7^\eff T_1~,\\
B&=&(1+\mhat_{K^*})\left[C_9 A_1+\frac{2\mhat_b}{\shat}(1-\mhat_{K^*})C_7^\eff T_2\right]~,\\ 
C&=&\frac{1}{1-\mhat_{K^*}^2}\left[(1-\mhat_{K^*})C_9 A_2
	+2\mhat_b C_7^\eff \left(T_3+\frac{1-\mhat_{K^*}^2}{\shat}T_2\right)\right]~,\\
D&=&\frac{1}{\shat}\left\{C_9\left[(1+\mhat_{K^*})A_1-(1-\mhat_{K^*})A_2
	-2\mhat_{K^*}A_0\right]-2\mhat_b C_7^\eff T_3\right\}~,\\
E&=&\frac{2}{1+\mhat_{K^*}}C_{10} V ~,\\
F&=&(1+\mhat_{K^*})C_{10} A_1 ~,\\
G&=&\frac{1}{1+\mhat_{K^*}}C_{10} A_2 ~,\\
H&=&\frac{1}{\shat}C_{10}\left[(1+\mhat_{K^*})A_1-(1-\mhat_{K^*})A_2-2\mhat_{K^*} A_0\right] ~.
\end{eqnarray}
%
%
%
%
\par
Another relevant process of $b\to s\ell^+\ell^-$ transition is $B_s\to\tau^+\tau^-$. 
The branching ratio is simply given by
\begin{equation}
\Br(B_s\to\tau^+\tau^-)
= \Br(B_s\to\tau^+\tau^-)_\SM \left|1
	+\frac{C_{10}^{\NP}-C_{10}'}{C_{10}^{\SM}}\right|^2~,
\label{Br_th}
\end{equation}
where $C_{10}^{\SM}$ is the SM value of $C_{10}$,
$C_{10}^\SM =-4.41$ \cite{DAlise2403,Damir1205}. 
%
%
%
\par
Now we move to the dineutrino decay modes.
The effective Hamiltonian for $b\to s\nu\nubar$ is 
\begin{equation}
\calH_\eff = -\frac{4G_F}{\sqrt{2}}V_{tb}V_{ts}^*\left(\frac{\alpha_\em}{4\pi}\right)
\Big[ C_L^\nu \calO_L^\nu + C_R^\nu \calO_R^\nu \Big]~,
\end{equation}
where
\begin{equation}
\calO_{L(R)}^\nu = 
\big[ \sbar\gamma^\mu P_{L(R)} b\big] 
\big[ \nubar\gamma_\mu(1-\gamma_5)\nu\big]~,
\end{equation}
and $C_L^\nu = C_{L,\SM}^\nu + C_{L,\LQ}^\nu$.
The scalar leptoquarks contribute to the Wilson coefficients as \cite{Becirevic1503,Crivellin1912}
\begin{eqnarray}
C_{L,\LQ}^\nu &=& \frac{1}{2\sqrt{2} G_F V_{tb} V_{ts}^*}\left(\frac{\pi}{\alpha_\em}\right)
\left[ \frac{x^{L*}_{23} x^L_{33}}{M_{S_1}^2} + \frac{z^*_{23} z_{33}}{M_{S_3}^2} \right]~,\nn\\
C_R^\nu          &=& -\frac{1}{2\sqrt{2} G_F V_{tb} V_{ts}^*}\left(\frac{\pi}{\alpha_\em}\right)
 \frac{y_{23} y^*_{33}}{M_{\tR2}^2}~. 
\end{eqnarray}
Note that we have $C_R^\nu = C_9^\prime$ in our framework.
\par
Numerical results for the branching ratios for $B^+\to K^+\nu\nubar$ and $B^0\to K^{*0}\nu\nubar$ 
are \cite{Chen2401} ,
\begin{eqnarray}
\Br(B^+\to K^+\nu\nubar) &=& 
3.46\times 10^{-8} \sum_\ell \left| C_L^{\nu_\ell}+C_R^{\nu_\ell}\right|^2~,\\
\Br(B^0\to K^{*0}\nu\nubar) &=& 
   6.84\times 10^{-8} \sum_\ell \left| C_L^{\nu_\ell}-C_R^{\nu_\ell}\right|^2 
+ 1.36\times 10^{-8} \sum_\ell \left| C_L^{\nu_\ell}+C_R^{\nu_\ell}\right|^2~.
\label{BrKsnunu}
\end{eqnarray}
%
%
%
\par
Now let us consider the $B_s$-$\Bbar_s$ mixing.
The effective Hamiltonian is
\begin{equation}
\calO_\eff = 
   C_1 (\bbar\gamma^\mu P_L s)(\bbar\gamma_\mu P_L s)
+ C_6(\bbar\gamma^\mu P_R s)(\bbar\gamma_\mu P_R s)~.
\end{equation}
The Wilson coefficients are given by
\begin{equation}
C_1 = C_1^\SM + C_1^{S_1} + C_1^{S_3} + C_1^{S_1+S_3}~,
\end{equation}
where \cite{Chen1703,Crivellin1912}
\begin{eqnarray}
C_1^{S_1} &=& \frac{1}{128\pi^2}\frac{1}{M_{S_1}^2}(x^{L*}_{23} x^L_{33})^2~,\nn\\
C_1^{S_3} &=& \frac{1}{128\pi^2}\frac{5}{M_{S_3}^2}(z^{*}_{23} z_{33})^2~,\nn\\
C_1^{S_1+S_3} &=& \frac{1}{128\pi^2}
                                 \frac{2}{M_{S_1}^2-M_{S_3}^2} 
                                 \log\left( \frac{M_{S_1}^2}{M_{S_3}^2} \right)
                                 (x^{L*}_{23} x^L_{33})(z^*_{23} z_{33})~,\nn\\
\end{eqnarray}
and \cite{Chen2205}
\begin{equation}
C_6^{\tR2} = \frac{1}{128\pi^2}\frac{2}{M_{\tR_2}^2}(y^*_{23} y_{33})^2~.
\end{equation}
The mass difference is given by
\begin{equation}
\Delta m_{B_s}=\Delta m_{B_s}^\SM\Bigg|
1 + \frac{\eta_B(C_1^{S_1}+C_1^{S_3}+C_1^{S_1+S_3})+\eta_1C_6^{\tR_2}}{\eta_B C_1^\SM}\Bigg|~,
\end{equation}
where $\eta_B$ and $\eta_1$ are the QCD corrections.
Note that $C_1^{S_1}$, $C_1^{S_3}$, and $C_6^{\tR_2}$ are positive definite for real couplings.
For simplicity we assume all the couplings are real in what follows.
In this case $y_{ij}$ appears only in the form of $(y_{23} y_{33})$ in  
$C_9^\prime$, $C_R^\nu$, and $C_6^{\tR_2}$,
thus it can be treated as a single coupling in our analysis.
%
%
%
%
%
\section{Results and discussions}
%
%
%
%
%
We try to fit the experimental data of Tables \ref{T_RDs}-\ref{T_b2s} by minimizing $\chi^2$ with scalar LQs
$S_1$, $\tR_2$, and $S_3$.
Table \ref{T_RDs} shows the experimental data for $B\to\Ds\ell\nubar$ observables, 
$R(\Ds)$, $P_\tau(D^*)$ and $F_L(D^*)$.
Other observables of $b\to c\ell\nubar$, $R(J/\Psi)$ and $R(\Lambda_c)$ are given in Table \ref{T_RJRLc}.
In Table \ref{T_b2s} observables of $b\to s $ transition, $\Delta m_{B_s}$ and $\Br(B^+ \to K^+\nu\nubar)$
are provided.
\par
To fit the data we scan the parameters $x_{23, 33}^L$, $x_{23}^R$, $y_{23}y_{33}$, $z_{23, 33}$, and 
$M_{S_1, \tR_2, S_3}$.
The scan ranges are $[-1, 1]$ for $x_{23, 33}^L$, $x_{23}^R$, $y_{23}y_{33}$, $z_{23, 33}$, and
$[1, 3]$ (TeV) for $M_{S_1, \tR_2, S_3}$.
%
%
%
\begin{table}
\begin{tabular}{c|cccccc}\hline\hline
                & $x^L_{23}$   &  $x^L_{33}$   & $x^R_{23}$   &  $y_{23}y_{33}$   &  $z_{23}$  \\ \hline
 best fits  & $0.12$         &  $-0.33$      & $0.63$         &  $0.03$              &  $0.04$   \\ \hline\hline
                &  $z_{33}$   & $M_{S_1}$(TeV) &  $M_{\tR_2}$(TeV) &  $M_{S_3}$(TeV)  &  $~~~\chi_\min^2/\dof~~~$ \\ \hline
 best fits  & $~~0.330$  &  $~~~~~1.29~~~~~$   &  $~~~~~1.07~~~~~$   &  $~~~~~2.62~~~~~$  & $1.63$ \\
\hline\hline
 \end{tabular}
 \caption{\label{T_best_fit} Best-fit values for fitting parameters.}
 \end{table} 
%
Our best-fit values of the fitting parameters are summarized in Table \ref{T_best_fit}.
%
%
%
%
%
We have $\chi^2_\min/\dof = 29.42/18 \simeq 1.63$, while for the SM we find
$\chi^2_\SM/\dof = 51.04/27\simeq 1.89$.
%
%
%
%
%
%
\begin{table}
\begin{tabular}{c|cc}\hline\hline
                                                                   &   best fits                  &   $1\sigma$ ranges \\ \hline
$C_{V_L}\times 10^2~~~$                  &  $~~~ - 0.83 ~~~$    &   $~~~ [-2.61, 3.86] ~~~$ \\
$C_{S_L}\times 10^2~~~$                         &  $~~~ 9.24 ~~~$    &  $~~~ [-12.14, 22.10] ~~~$ \\
$C_T     \times 10^2~~~$                           &  $~~~ -3.13 ~~~$   &  $~~~ [-5.00, 0.53] ~~~$ \\
$C_9^\LQ  ~~~ $                                        &  $~~~ 1.16 ~~~$      &  $~~~ [-23.25, 20.35]~~~$ \\       
$C_9^\prime ~~~ $                                     &  $~~~ -8.37 ~~~$     &  $~~~ [-15.66, 14.72]~~~$ \\   
$C_{L,\LQ}^\nu  ~~~ $                                &  $~~~ -6.59 ~~~$    &  $~~~ [-9.28, 20.56]~~~$ \\   
$C_R^\nu ~~~ $                                          &  $~~~ -8.37 ~~~$     &  $~~~ [-15.66, 14.72]~~~$ \\   
$(C_1^{S_1}/C_1^\SM)\times 10^2$           &  $~~~ 0.82 ~~~$      &  $~~~ [1.09\times 10^{-9}, 9.81] ~~~$ \\
$(C_1^{S_3}/C_1^\SM)\times 10^2$           &  $~~~ 0.11 ~~~$      &  $~~~ [4.08\times 10^{-9}, 8.66] ~~~$ \\
$(C_1^{S_1+S_3}/C_1^\SM)\times 10^2$   &  $~~~ -0.25 ~~~$      &  $~~~ [-5.58, 2.06] ~~~$ \\
$(C_6^{\tR2}/C_1^\SM)\times 10^2$           &  $~~~ 1.53 ~~~$      &  $~~~ [2.62\times 10^{-8}, 8.34] ~~~$ \\
 \hline\hline
 \end{tabular}
 \caption{\label{T_BF_WC} Best-fit values for Wilson coefficients and allowed ranges at $1\sigma$.}
 \end{table} 
%
In Table \ref{T_BF_WC} best-fit values and allowed regions at $1\sigma$ 
for the relevant Wilson coefficients are provided.
The best-fit value of $C_{V_L}$ is negative while
that of $C_{S_L}$ is positive and larger compared to $|C_{V_L}|$.
Thus in our fitting the $R(\Ds)$ anomaly is explained dominantly by $C_{S_L}$.
This is quite different feature from other researches \cite{DAlise2403,Iguro2405,JPL2411}.
In Eq.\ (\ref{E_CVL}), the leading contribution comes from
$C_{V_L}\sim \left(x_{23}^{L} x_{33}^L/M_{S_1}^2 - z_{23} z_{33}/M_{S_3}^2\right)$. 
Note that $x_{23}^{L} x_{33}^L$ and $z_{23} z_{33}$ also appear in 
$C_{L,\LQ}^\nu\sim \left(x_{23}^{L} x_{33}^L/M_{S_1}^2 + z_{23} z_{33}/M_{S_3}^2\right)$ for $b\to s\nu\nubar$
and $C_1^{S_1+S_3}\sim \left(x_{23}^{L} x_{33}^L\right) \left(z_{23} z_{33}\right)$ for $\Delta m_{B_s}$.
To fit the experimental data of $\Delta m_{B_s} < \Delta m_{B_s}^\SM$, 
negative values of $C_1^{S_1+S_3}$ are preferred because other contributions are constructive.
It means that $\left(x_{23}^L x_{33}^L\right)$ and $\left(z_{23} z_{33}\right)$ tend to have different signs.
Our best fits prefer that $C_9^\LQ\sim z_{23} z_{33} >0 $ and thus 
$C_{V_L}\sim \left(x_{23}^{L} x_{33}^L/M_{S_1}^2 - z_{23} z_{33}/M_{S_3}^2\right) < 0 $ such that
$C_{L,\LQ}^\nu\sim \left(x_{23}^{L} x_{33}^L/M_{S_1}^2 + z_{23} z_{33}/M_{S_3}^2\right) < 0 $.
The result of $C_{L,\LQ}^{\nu,\best} <0 $ is also consistent with our previous work of \cite{JPL2509}.
It should be noted that a region of opposite signs is also possible at the $1\sigma$ level.
\par
The allowed ranges for $C_9^\LQ$, $C_9^\prime=C_R^\nu$, and $C_{L,\LQ}^\nu$ are rather wide
because they are not constrained strongly by experiment yet.
Other researches for the muonic sector suggest that $C_9^{\ell =\mu,\LQ}\approx -1$ \cite{Crivellin2505}.
If the LFU holds, $C_9^\LQ=C_9^{\ell=\mu,\LQ}\equiv C_9^U$ must be $\approx -1$, but our best fit is opposite.
The value of $C_9^\LQ=-1$ is still within the allowed range as shown in Table \ref{T_BF_WC},
but our best-fit value prefers the LFU violating couplings.
So the precise determination of $C_9$ in tauonic sector is very important to check the LFU violation.
%
%
%
\begin{table}
\begin{tabular}{c|ccc}\hline\hline
                                               & SM predictions          &   best fits                &   $1\sigma$ ranges \\ \hline
$R(D) ~~~ $                           & $0.290\pm 0.003$     &  $~~~ 0.33 ~~~$    &   $~~~ [0.24, 0.41] ~~~$ \\
$R(D^*) ~~~$                         & $0.248\pm 0.001$      &  $~~~ 0.29 ~~~$    &  $~~~ [0.25, 0.33] ~~~$ \\
$P_\tau(D) ~~~$                    & $0.331\pm 0.004$      &  $~~~ 0.44 ~~~$   &  $~~~ [0.21, 0.53] ~~~$ \\
$P_\tau(D^*)  ~~~ $               & $-0.497\pm 0.007$     &  $~~~ -0.48 ~~~$  &  $~~~ [-0.53, -0.44]~~~$ \\       
$F_L(D^*) ~~~ $                    & $0.464\pm 0.003$      &  $~~~ 0.44 ~~~$    &  $~~~ [0.43, 0.47]~~~$ \\   
$R(J/\Psi) ~~~ $                     & $0.258\pm0.004$       & $~~~ 0.30 ~~~$    &  $~~~ [0.25, 0.33]~~~$ \\
$R(\Lambda_c) ~~~ $            & $0.324\pm 0.004$      & $~~~ 0.36 ~~~$    &  $~~~ [0.32, 0.40]~~~$ \\
$\Delta m_{B_s}\times 10^{-12}\cdot{\rm sec} ~~~ $  
                                         & $18.4^{+0.7}_{-1.2}$   &  $~~~ 18.80 ~~~$    &  $~~~ [18.40, 20.11]~~~$ \\   
$\Br(B^+\to K^+\nu\nubar)\times 10^6 ~~~ $        
                                          & $4.43\pm0.31$       &  $~~~ 18.43 ~~~$     &  $~~~ [2.76, 28.96]~~~$ \\   
$\Br(B^0\to K^{*0}\nu\nubar)\times 10^6 ~~~ $     
                                          & $9.47\pm1.40$       &  $~~~ 14.12 ~~~$     &  $~~~ [6.55, 18.00]~~~$ \\   
$\Br(B^+\to K^+\tau^+\tau^-)\times 10^7 ~~~ $     
                         & $1.17\pm 0.14$                        &  $~~~ 0.72 ~~~$     &  $~~~ [9.46\times 10^{-3}, 54.89]~~~$ \\   
$\Br(B^0\to K^{*0}\tau^+\tau^-)\times 10^7 ~~~ $  
                         & $0.87\pm0.11$                        &  $~~~ 12.40 ~~~$     &  $~~~ [1.32\times 10^{-2}, 52.58]~~~$ \\   
$\Br(B_s\to \tau^+\tau^-)\times 10^6 ~~~ $            
                         & $0.773\pm0.049$                     &  $~~~ 7.72 ~~~$     &  $~~~ [1.17\times 10^{-7}, 30.52]~~~$ \\   
$\Br(B_c^-\to \tau^-\nubar)\times 10^2 ~~~ $         
                        & $2.29\pm0.09$                         &  $~~~ 0.80 ~~~$     &  $~~~ [2.47\times 10^{-3}, 5. 43]~~~$ \\   
 \hline\hline
 \end{tabular}
 \caption{\label{T_BF_Obs} Best-fit values for experimental observables and allowed ranges at $1\sigma$.
 A kinematic cut of $q^2>15~\GeV^2$ is imposed in $\Br(B\to\Ks\tau^+\tau^-)$.}
 \end{table} 
%
\par
Table \ref{T_BF_Obs} shows best-fit values and allowed regions at $1\sigma$ for physical observables.
For $\Br(B\to \Ks\tau^+\tau^-)$ a kinematic cut of $q^2 > 15~\GeV^2$ is imposed
to remove $\Psi(2S)$ resonance and exclude long-distance charmonium pollutions.
.
Note that $\Br(B^0\to K^{*0}\nu\nubar)$, $\Br(B^+\to K^+\tau^+\tau^-)$, $\Br(B^0\to K^{*0}\tau^+\tau^-)$, 
$\Br(B_s\to\tau^+\tau^-)$, and $\Br(B_c^-\to\tau^-\nubar)$ are not observed yet and only upper bounds exist.
In our predictions the neutral modes of $\Br(B^0\to K^{*0}\nu\nubar)$, $\Br(B^0\to K^{*0}\tau^+\tau^-)$, 
and $\Br(B_s\to\tau^+\tau^-)$ are expected to be larger than the SM values, while the charged modes of 
$\Br(B^+\to K^+\tau^+\tau^-)$ and $\Br(B_c^-\to\tau^-\nubar)$ are smaller.
Especially $\Br(B_s\to\tau^+\tau^-)_\best$ is much larger than its corresponding SM values,
so the measurement of the branching ratio would be a good probe for NP. 
And our $1\sigma$ range of $\Br(B_c^-\to\tau^-\nubar)$ is well below the upper bound of $30\%$.
Actually it also satisfies more aggressive bound of $\Br(B_c^-\to\tau^-\nubar) < 0.1$.
%
\begin{figure}
\begin{tabular}{cc}
\hspace{-1cm}\includegraphics[scale=0.12]{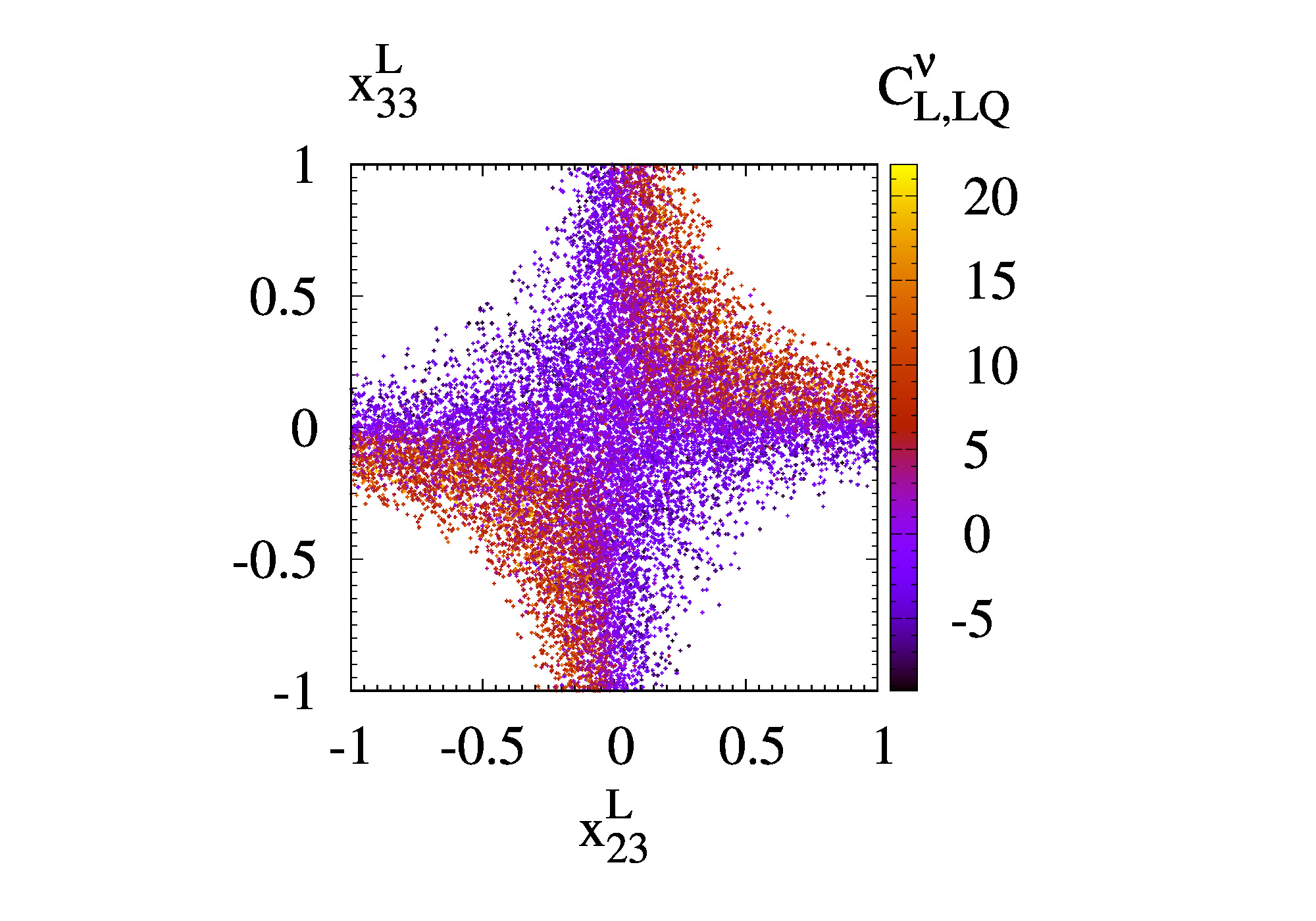}&
\hspace{-1cm}\includegraphics[scale=0.12]{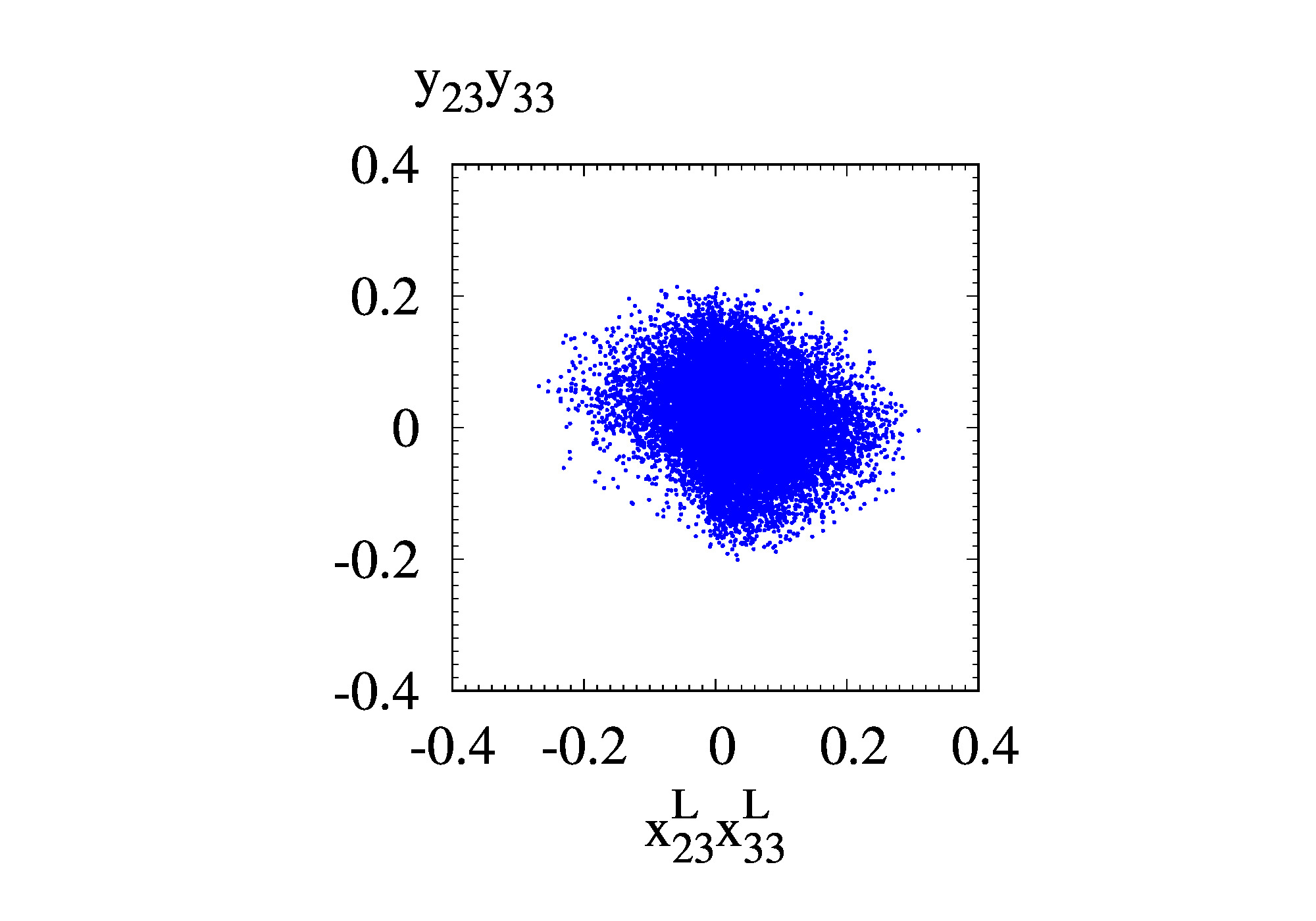}\\
(a) & (b)\\
\hspace{-1cm}\includegraphics[scale=0.12]{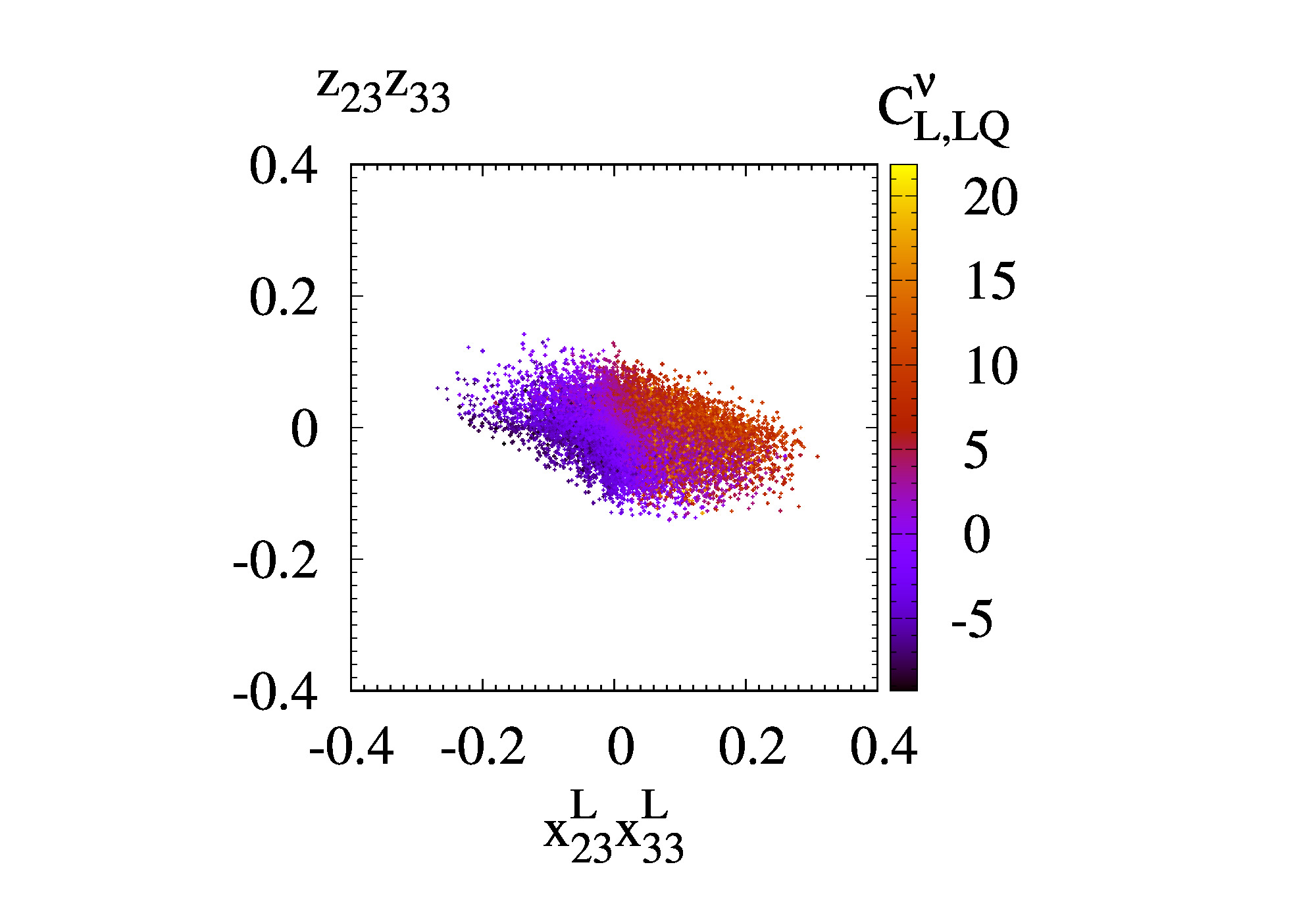}&
\hspace{-1cm}\includegraphics[scale=0.12]{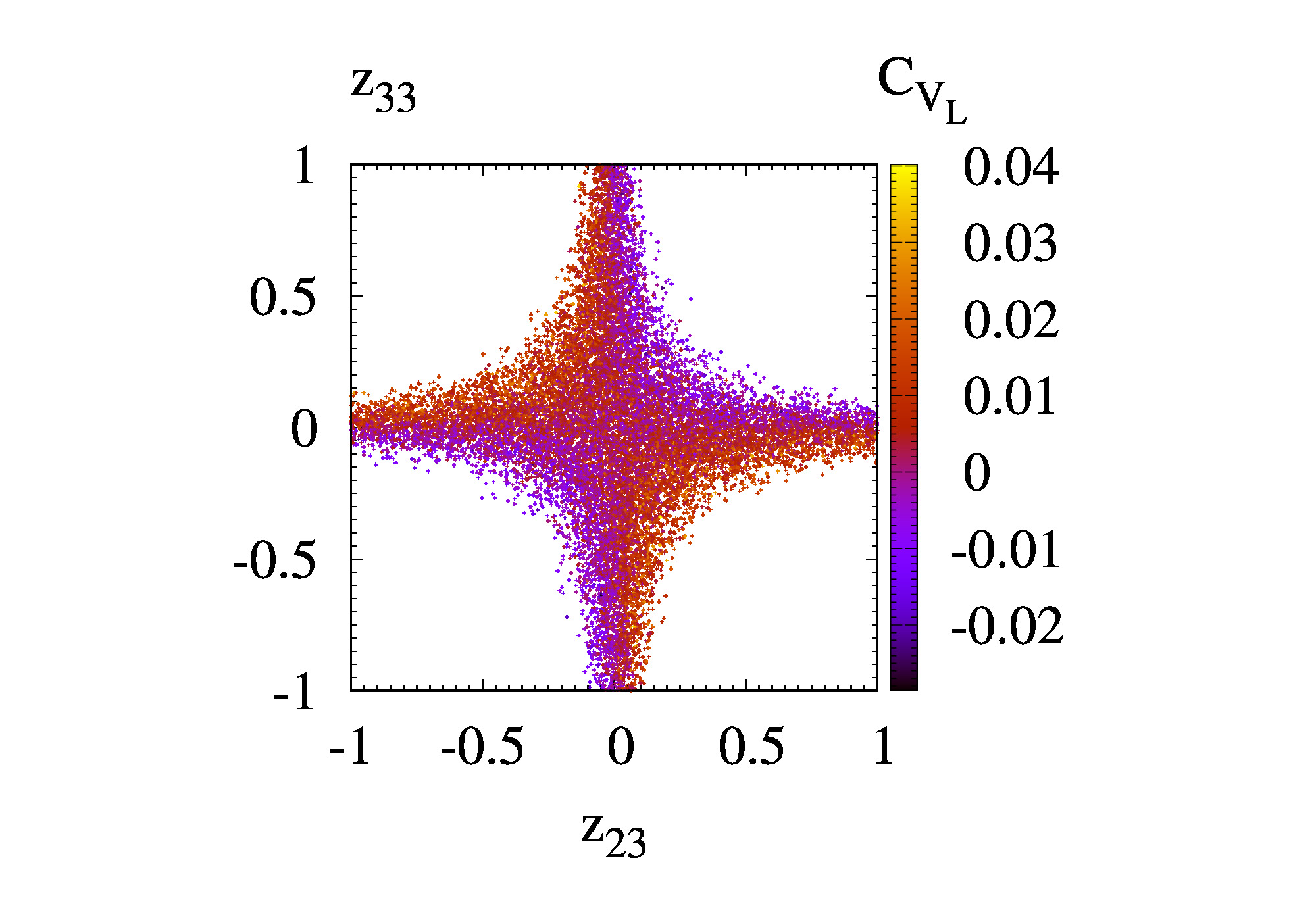}\\
(c) & (d)\\
\end{tabular}
\caption{\label{F_yz} Allowed regions at the $2\sigma$ level for
(a) $x_{33}^L$ vs. $x_{23}^L$ with respect to $C_{L,\LQ}^\nu$,  
(b) $y_{23} y_{33}$ vs. $x_{23}^L x_{33}^L$,  
(c) $z_{23} z_{33}$ vs. $x_{23}^L x_{33}^L$ with respect to $C_{L,\LQ}^\nu$, and
(d) $z_{33}$ vs. $z_{23}$ with respect to $C_{V_L}$
.}
\end{figure}
%
\par
Figure \ref{F_yz} plots allowed regions of various couplings.
As mentioned before $y_{23}y_{33}$ can be treated as a single coupling in our analysis.
Also for $x_{23}^L$ and $x_{33}^L$ or $z_{23}$ and $z_{33}$, 
products of $x_{23}^L x_{33}^L$ and $z_{23} z_{33}$ play important roles to describe the Wilson coefficients. 
They correspond to the fermionic coupling coefficients $A_j$ of \cite{JPL2110}.
As shown in Fig.\ \ref{F_yz} (c), the range of $\left|x_{23}^L x_{33}^L\right|$ is wider than 
that of $|z_{23} z_{33}|$.
It implies that in case of $C_1^{S_1+S_3}\sim\left(x_{23}^L x_{33}^L\right)\left(z_{23} z_{33}\right) < 0$,
negative values of $C_{L,\LQ}^\nu\sim\left(x_{23}^L x_{33}^L/M_{S_1}^2 + z_{23} z_{33}/M_{S_3}^2\right)$
prefer $x_{23}^L x_{33}^L < 0$ and $z_{23} z_{33} >0$ (noting that $M_{S_1}^\best < M_{S_3}^\best$).
The situation is represented in Figs.\ \ref{F_yz} (a) and (c) by colors of $C_{L,\LQ}^\nu$.
As discussed before, in this case $C_{V_L}$ tends to be negative.
Figure \ref{F_yz} (d) shows that positive $z_{23} z_{33}$ favors $C_{V_L} < 0$, as expected.
In our analysis the right-handed coupling $x_{23}^R$ appears only in $C_{S_L}=-4C_T$,
and thus it is constrained rather weakly compared with other couplings.
%
%
%
%
\begin{figure}
\begin{tabular}{cc}
\hspace{-1cm}\includegraphics[scale=0.12]{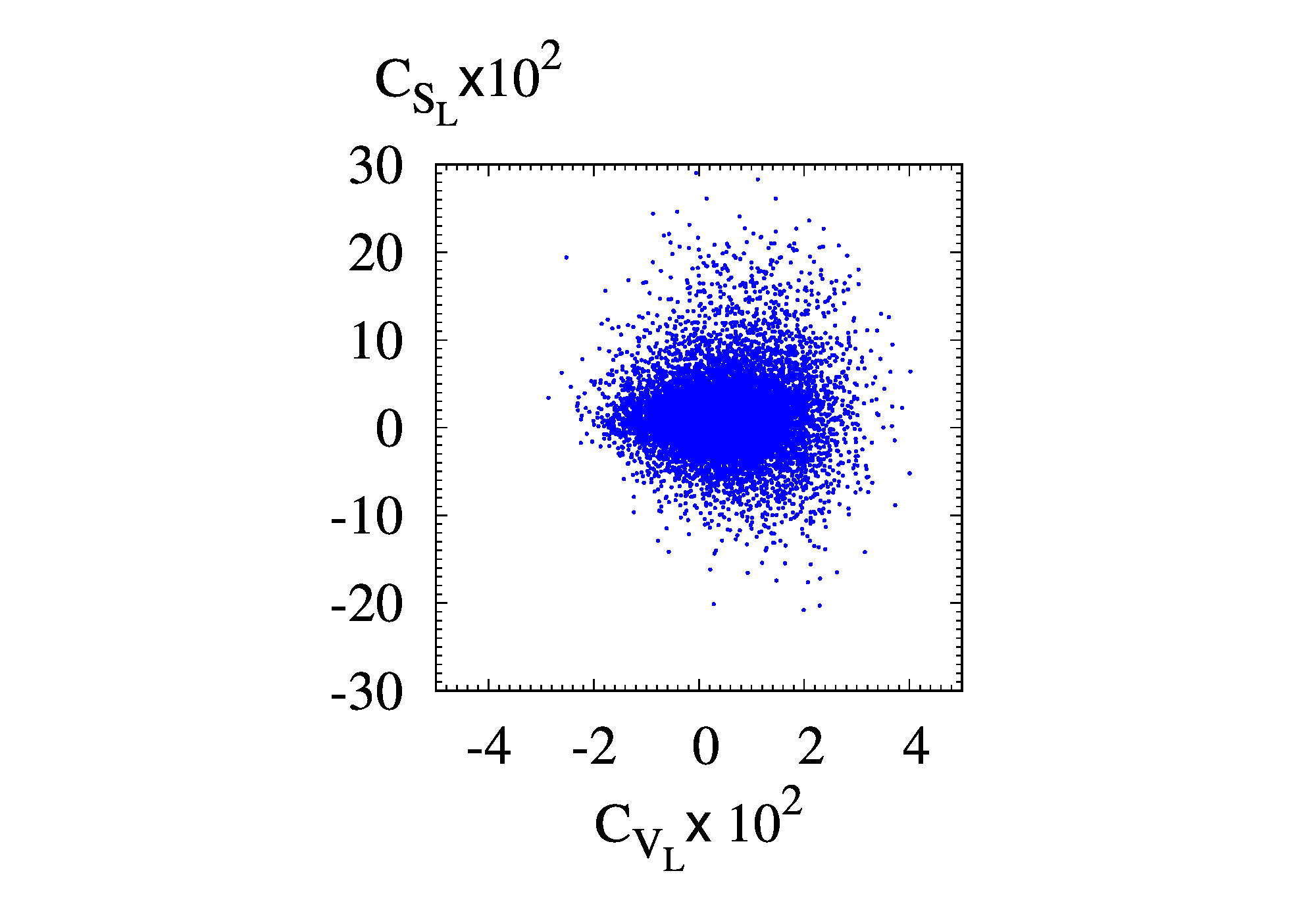}&
\hspace{-1cm}\includegraphics[scale=0.12]{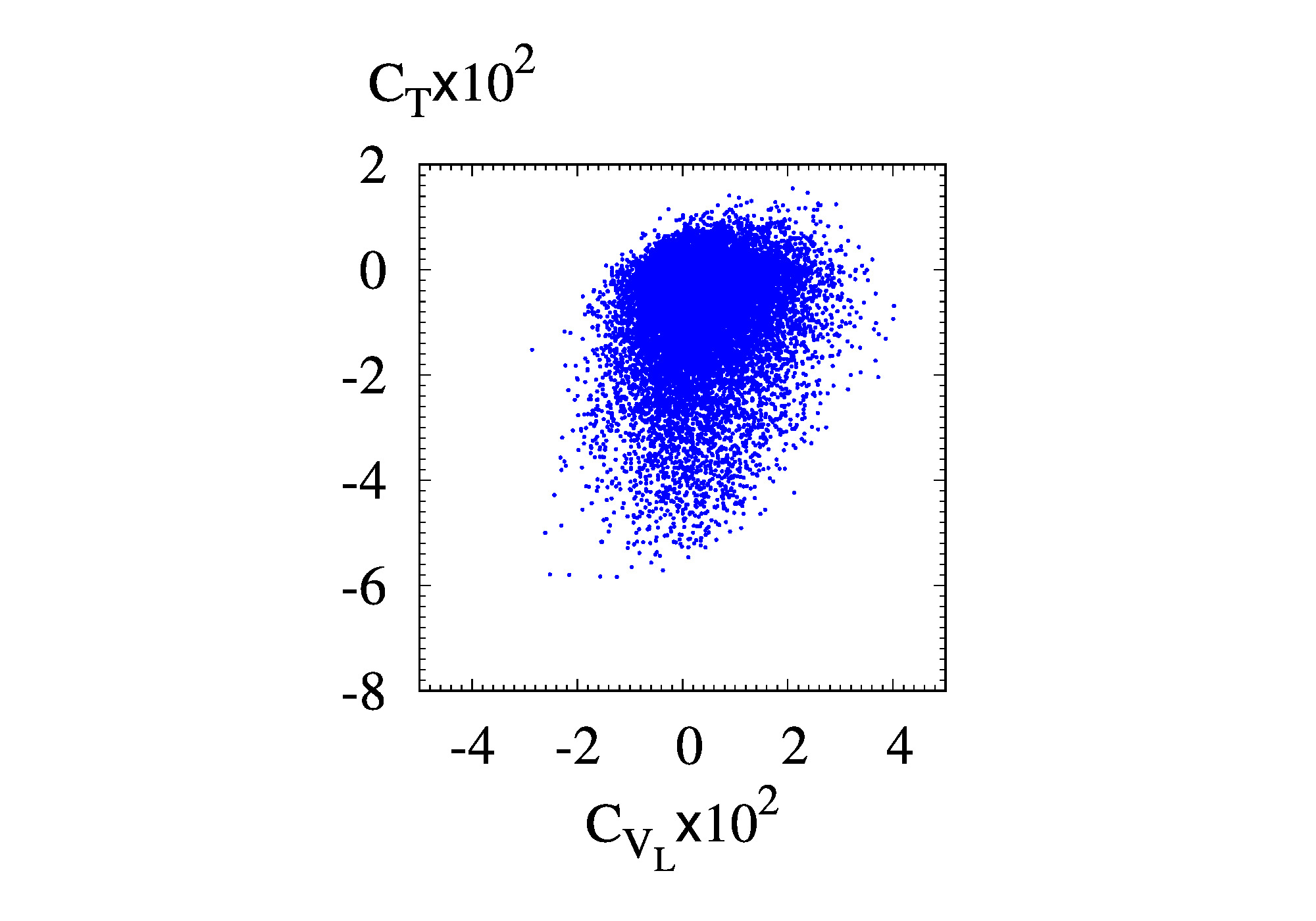}\\
(a) & (b)\\
\hspace{-1cm}\includegraphics[scale=0.12]{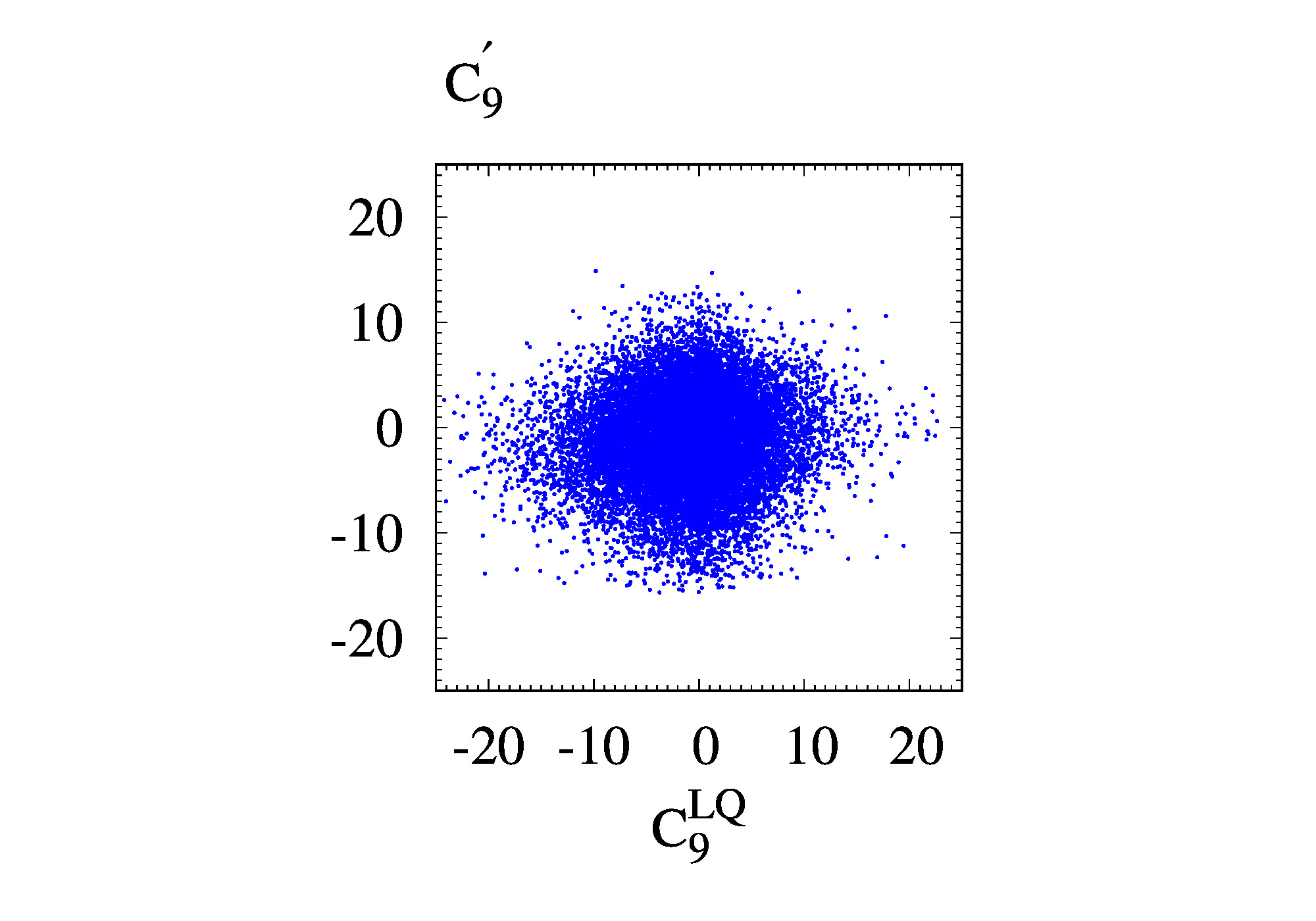}&
\hspace{-1cm}\includegraphics[scale=0.12]{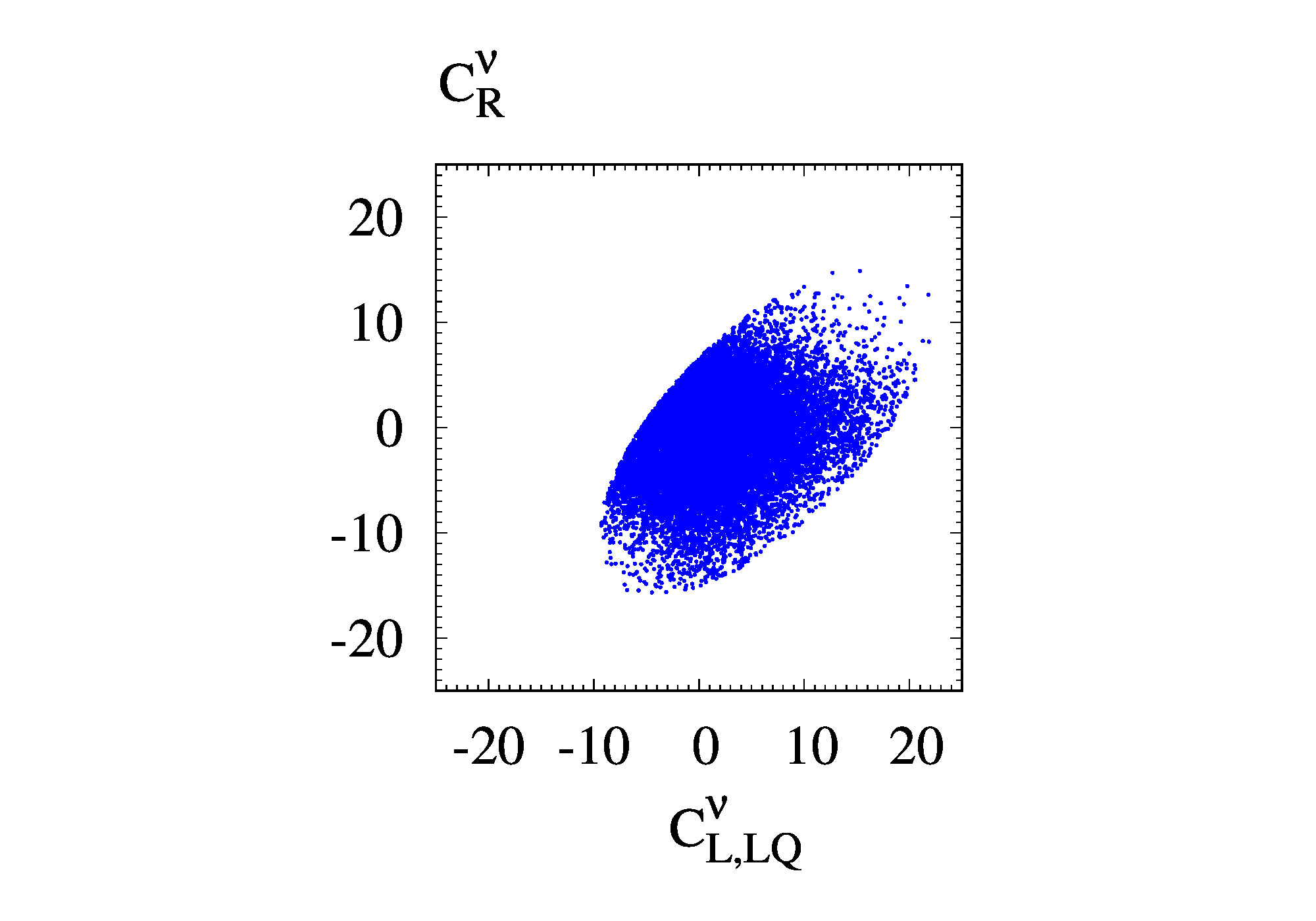}\\
(c) & (d)
\end{tabular}
\caption{\label{F_WC} Allowed regions at the $2\sigma$ level for
(a) $C_{S_L}$ vs. $C_{V_L}$,
(b) $C_T$ vs. $C_{V_L}$,
(c) $C_9^\prime$ vs. $C_9^\LQ$, and
(d) $C_R^\nu$ vs. $C_{L,\LQ}^\nu$.
}
\end{figure}
%
\begin{figure}
\begin{tabular}{cc}
\hspace{-1cm}\includegraphics[scale=0.12]{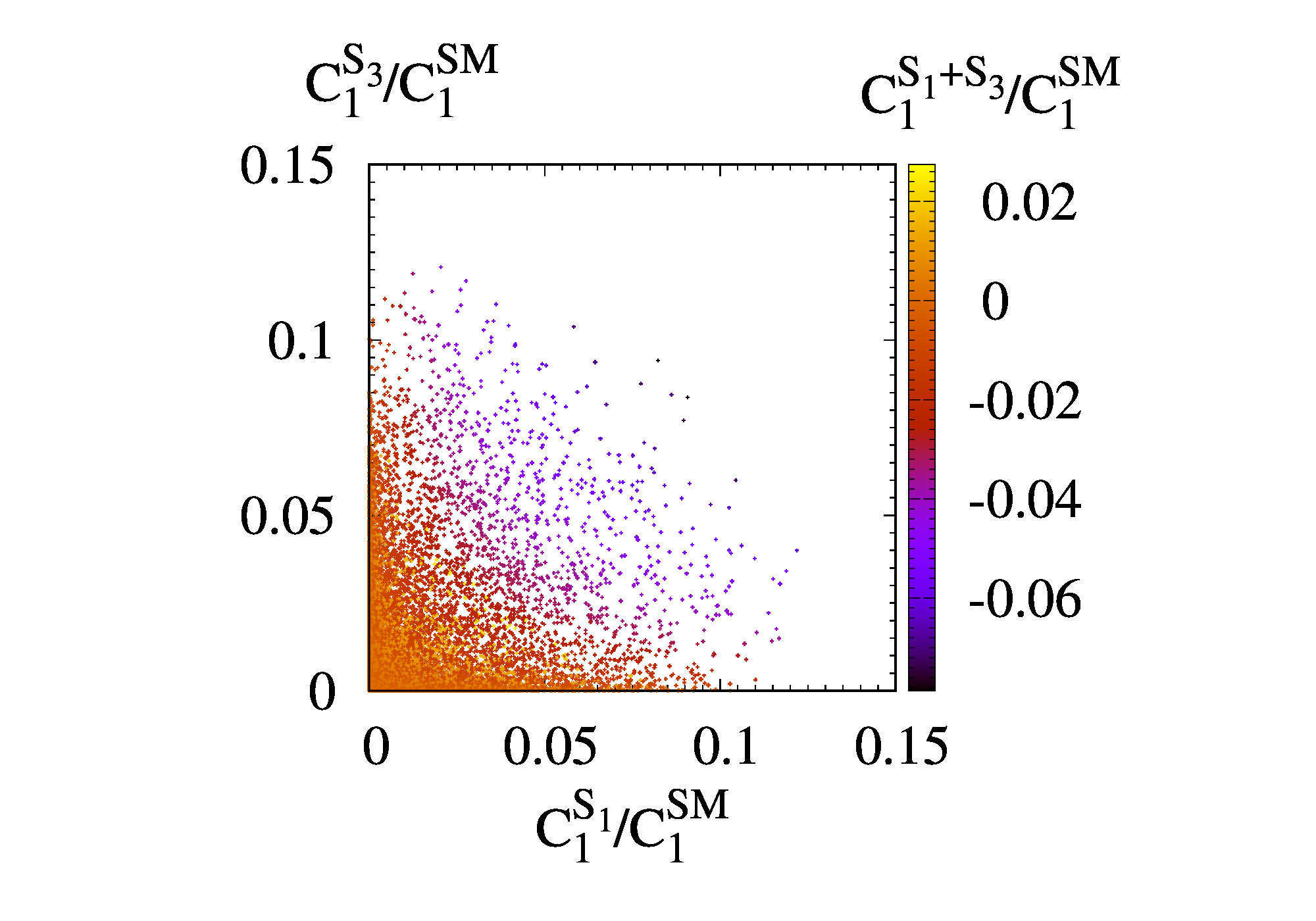} &
\hspace{-1cm}\includegraphics[scale=0.12]{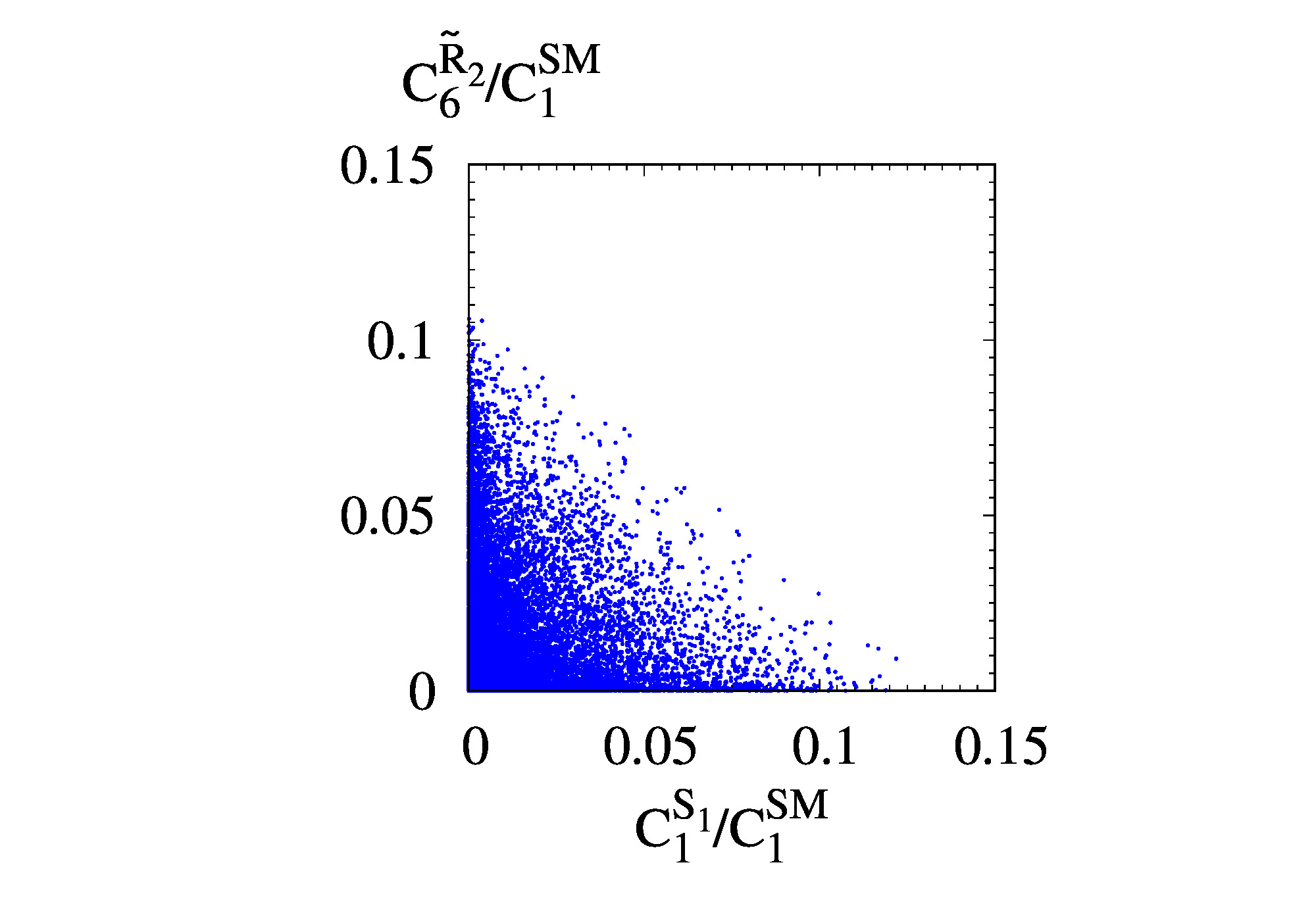}\\
(a) & (b) \\
\hspace{-1cm}\includegraphics[scale=0.12]{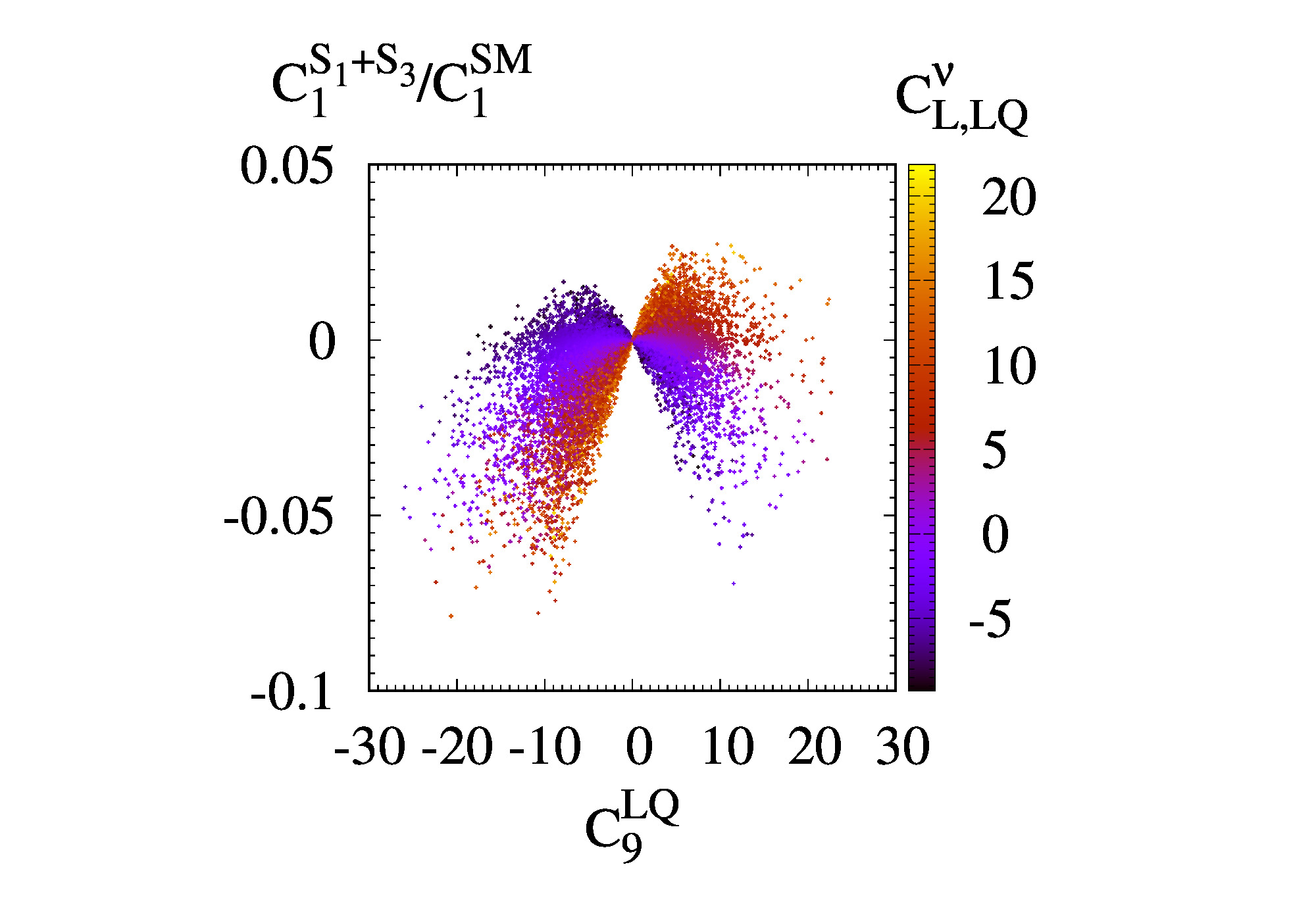} &
\hspace{-1cm}\includegraphics[scale=0.12]{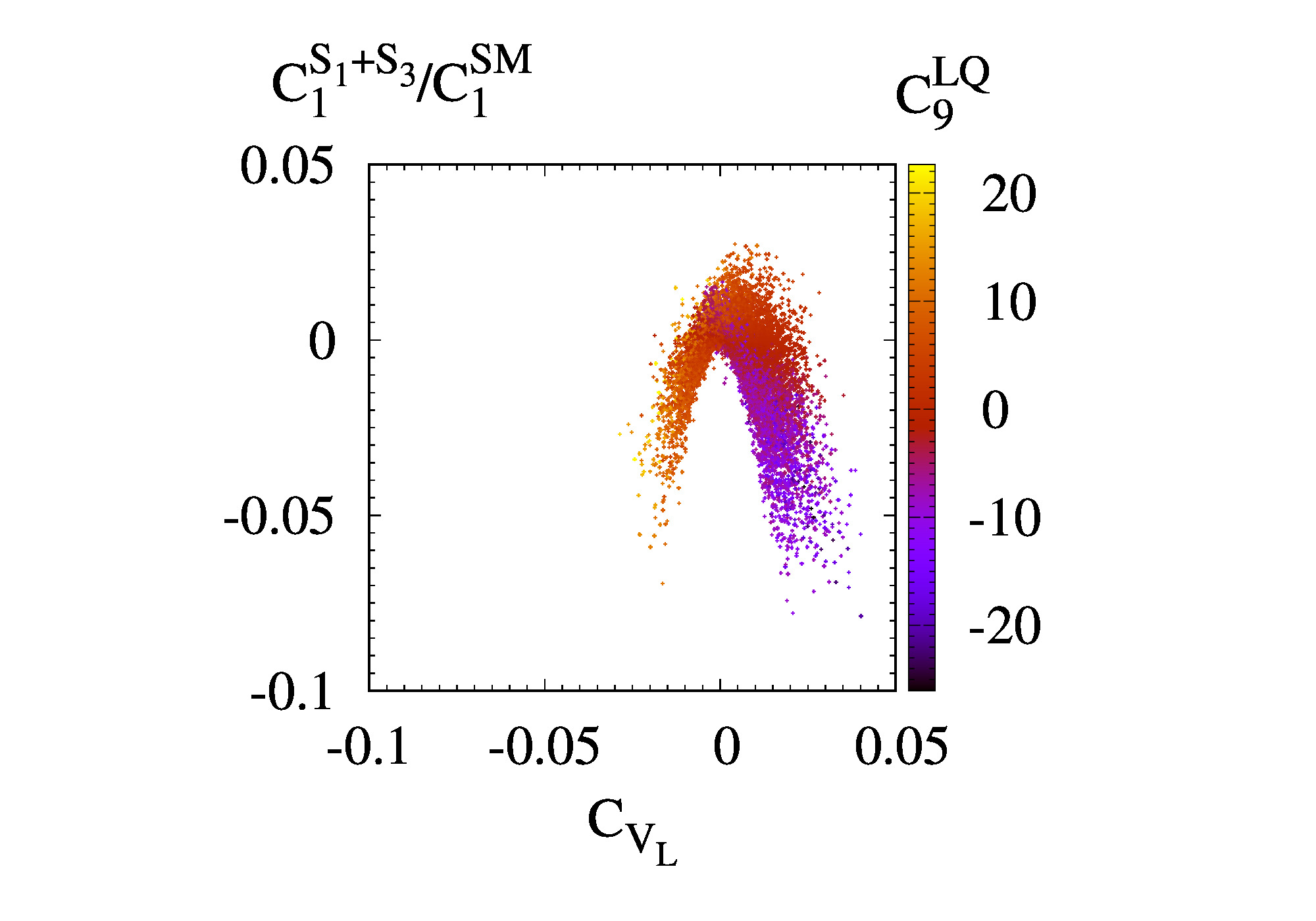}  \\
(c) & (d)
\end{tabular}
\caption{\label{F_WC2} Allowed regions at the $2\sigma$ level for
(a) $C_1^{S_3}/C_1^\SM$ vs. $C_1^{S_1}/C_1^\SM$ with respect to $C_1^{S_1+S_3}/C_1^\SM$, and
(b) $C_6^{\tR_2}/C_1^\SM$ vs. $C_1^{S_1}/C_1^\SM$, 
(c) $C_1^{S_1+S_3}/C_1^\SM$ vs. $C_9^\LQ$ with respect to $C_{L,\LQ}^\nu$, and
(d) $C_1^{S_1+S_3}/C_1^\SM$ vs. $C_{V_L}$ with respect to $C_9^\LQ$.
}
\end{figure}
%
\par
In Figs.\ \ref{F_WC} and \ref{F_WC2} we show the allowed regions of the relevant Wilson coefficients.
Figure \ref{F_WC}  (a) shows that in our fitting $C_{S_L}$ is dominant over $C_{V_L}$.
Our best-fit values are $C_{V_L}^\best < 0$ and $C_{S_L}^\best > 0$, 
but opposite signs are also allowed at the $2\sigma$ level.
As for $C_T$,  it favors negative values as shown in Fig.\ \ref{F_WC} (b).
Note that the scalar singlet $S_1$ requires $C_{S_L} = -4C_T$, as seen in Eq.\ (\ref{E_CSL}).
In previous works of \cite{JPL2411,JPL2502,JPL2509}, we parameterize the Wilson coefficients
in a generic way as $C_j\sim A_j (v/M_\NP)^\alpha$.
In \cite{JPL2411} we found that nonzero $C_T$ is very important in regarding $P_\tau(D)$ where
it contains a term $+0.17|C_T|^2 -1.09(1+C_{V_L})C_T$.
For positive $C_T$, $P_\tau(D)$ can be negative.
In our case $C_T$ is mostly negative and is expected to enhance $P_\tau(D)$.
As will be seen later, we have $P_\tau(D) > 0$ in this analysis.
\par
When LFU holds, $C_9^\LQ$ and $C_9^\prime$ would be $\approx -1$ \cite{Geng1704}.
As mentioned before, our best fits are $C_9^{\LQ,\best} = +1.16$ and $C_9^{\prime,\best} = -8.37$.
But Fig.\ \ref{F_WC} shows that different signs of $C_9^{\LQ,\prime}$ are also possible. 
To pin down the ranges more measurements on $b\to s\tau^+\tau^-$ are needed.
Figure \ref{F_WC} (d) depicts $C_R^\nu$ vs. $C_{L,\LQ}^\nu$.
It must be compared with previous work of \cite{JPL2509} (see Fig.\ 1 (e) therein)
where $C_{L,\NP}^{\nu_e}=C_{L,\NP}^{\nu_\mu}\ne 0$.
In both cases the best-fit values of $C_R^\nu$ and $C_{L,\LQ}^\nu$ are negatively order one.
\par
Figure \ref{F_WC2} plots the Wilson coefficients relevant for $\Delta m_{B_s}$.
Figures \ref{F_WC2} (c) and (d) represent graphically our discussions 
about the signs of $C_9^\LQ$ and $C_{V_L}$.
Note that $C_1^{S_1+S_3}$ and $C_{V_L}$ both involve $x_{23}^{L} x_{33}^L$ and $z_{23} z_{33}$.
As one can see in Fig.\ \ref{F_WC2} (c), for $C_1^{S_1+S_3} < 0$ and $C_{L,\LQ}^\nu < 0$,
$C_9^\LQ$ can have both positive and negative values. 
But for negatively large $C_1^{S_1+S_3}$, $C_9^\LQ > 0$ is favored.
In this case, Fig.\ \ref{F_WC2} (d) tells that negative $C_{V_L}$ is preferred. 
But if the situation is reversed positive $C_{V_L}$ is also possible.
%
%
%
%
%
\begin{figure}
\begin{tabular}{cc}
\hspace{-1cm}\includegraphics[scale=0.12]{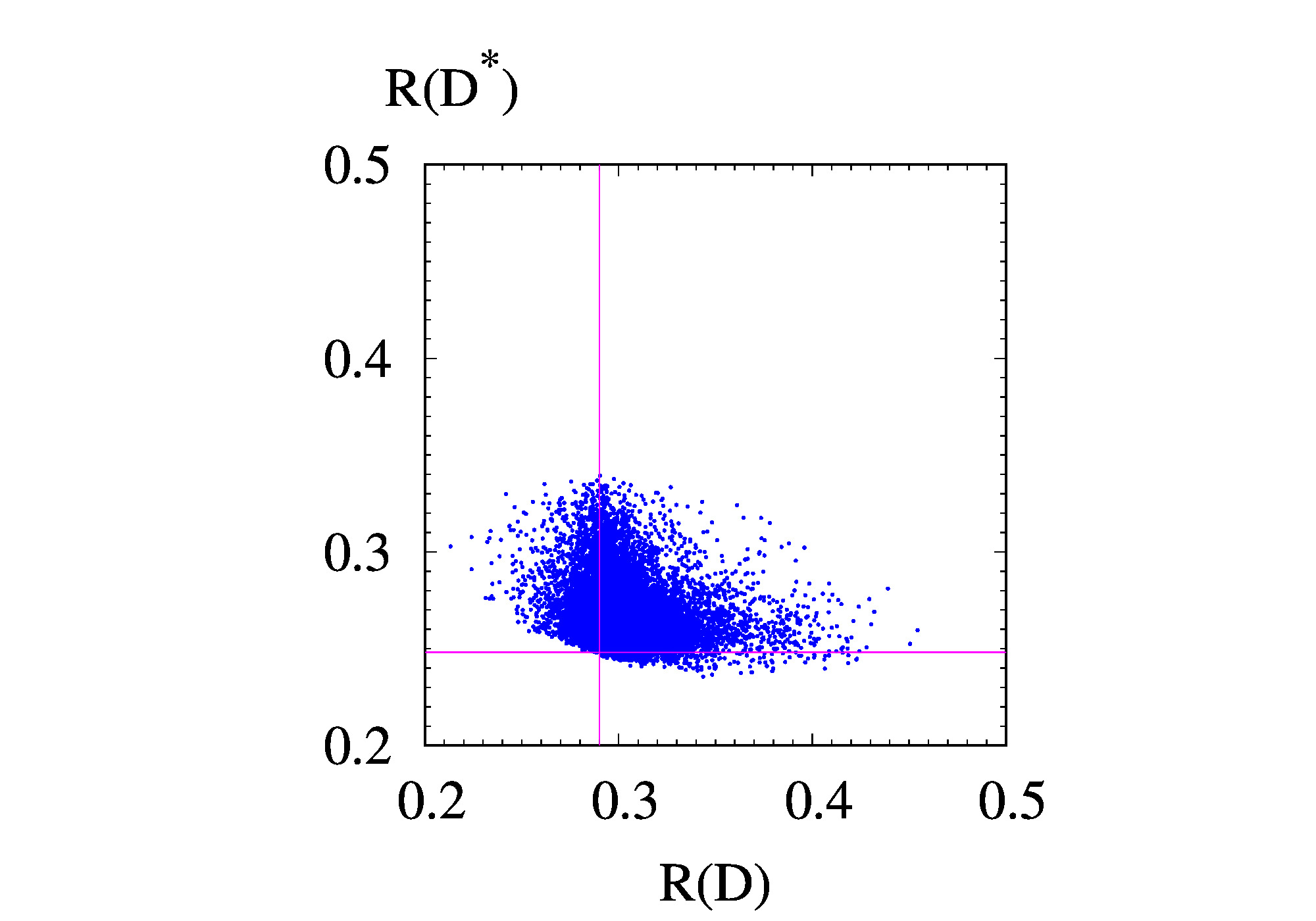}&
\hspace{-1cm}\includegraphics[scale=0.12]{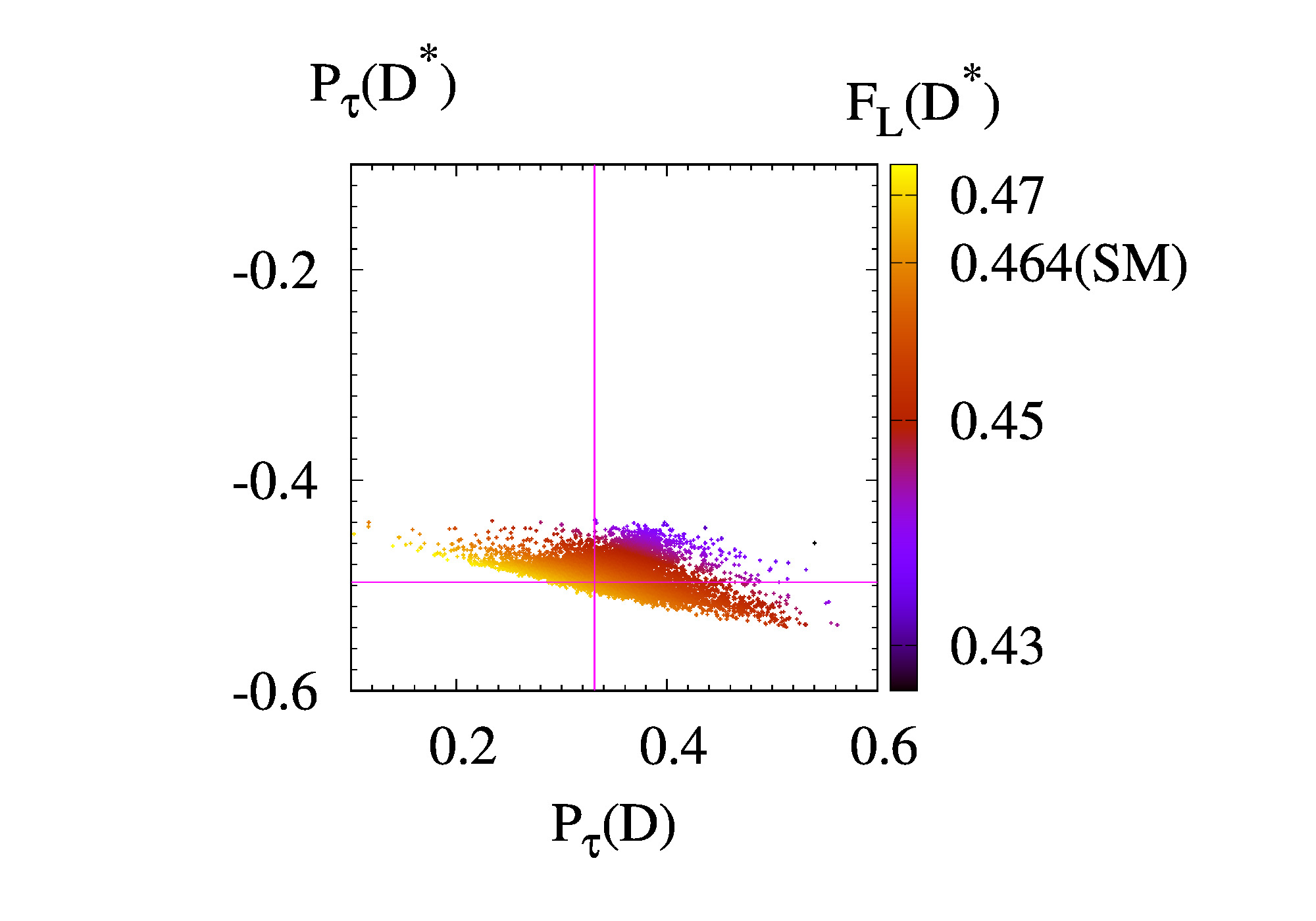}\\
(a) & (b) \\
\hspace{-1cm}\includegraphics[scale=0.12]{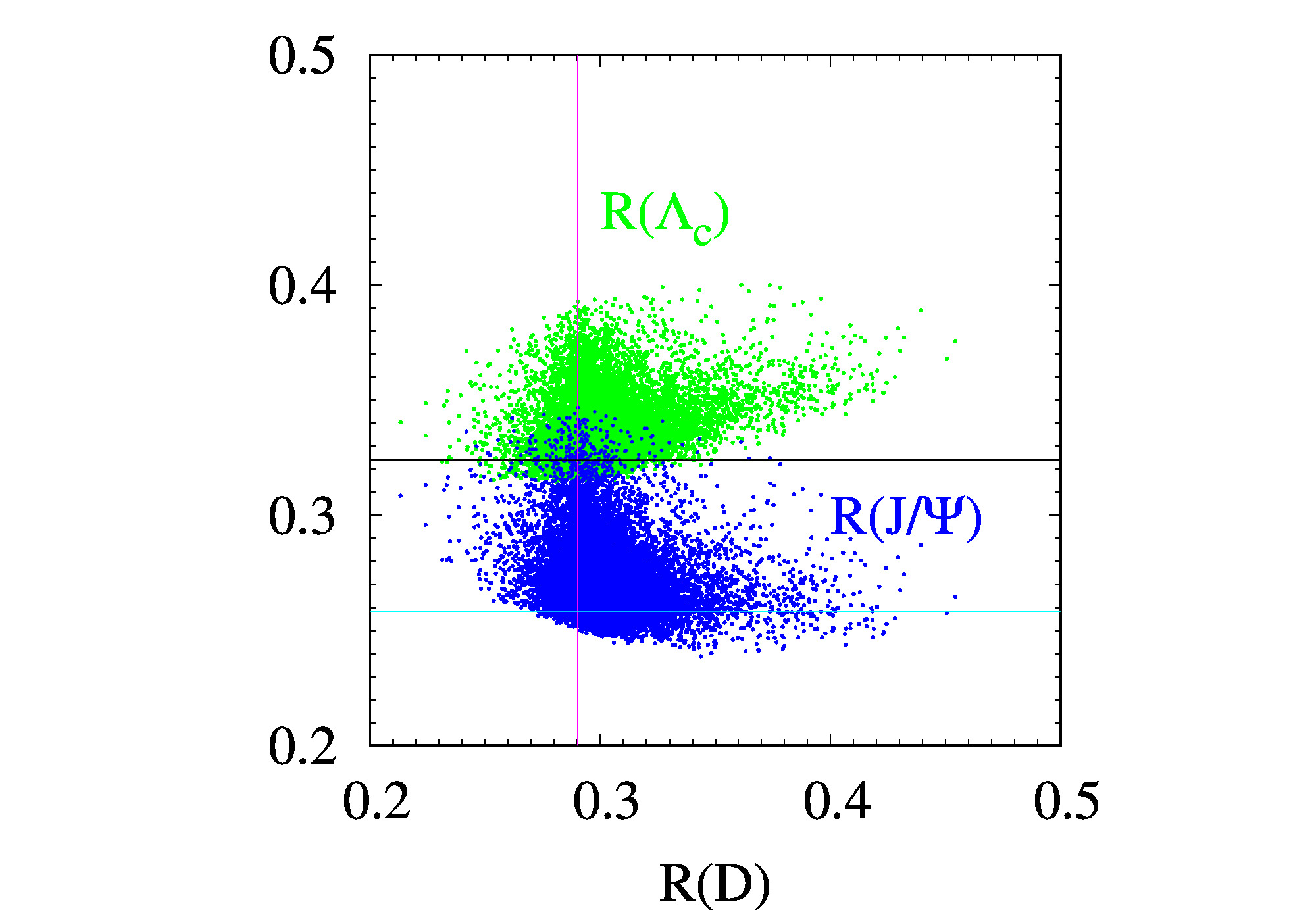}& 
\hspace{-1cm}\includegraphics[scale=0.12]{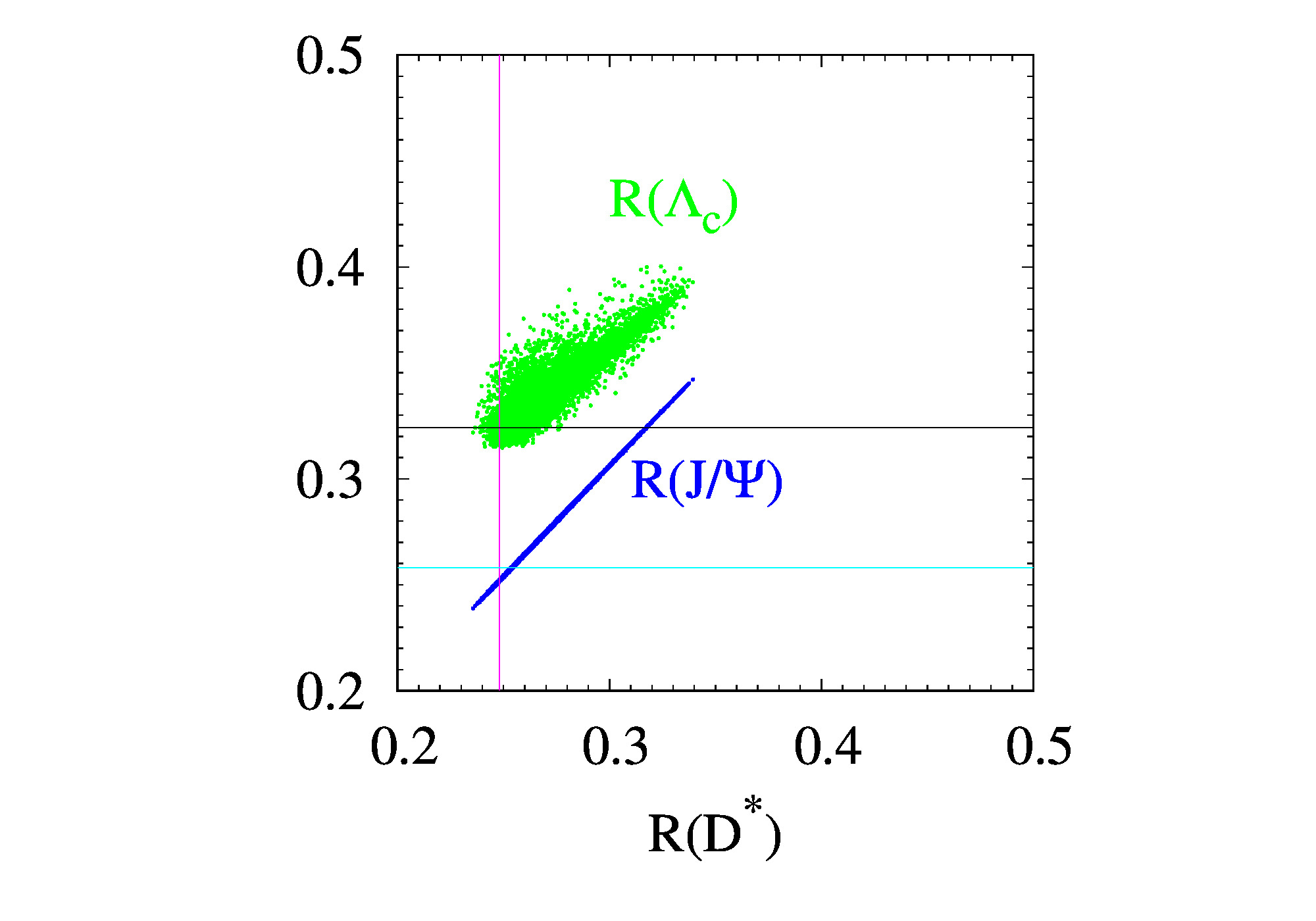}\\
(c) & (d) \\
\hspace{-1cm}\includegraphics[scale=0.12]{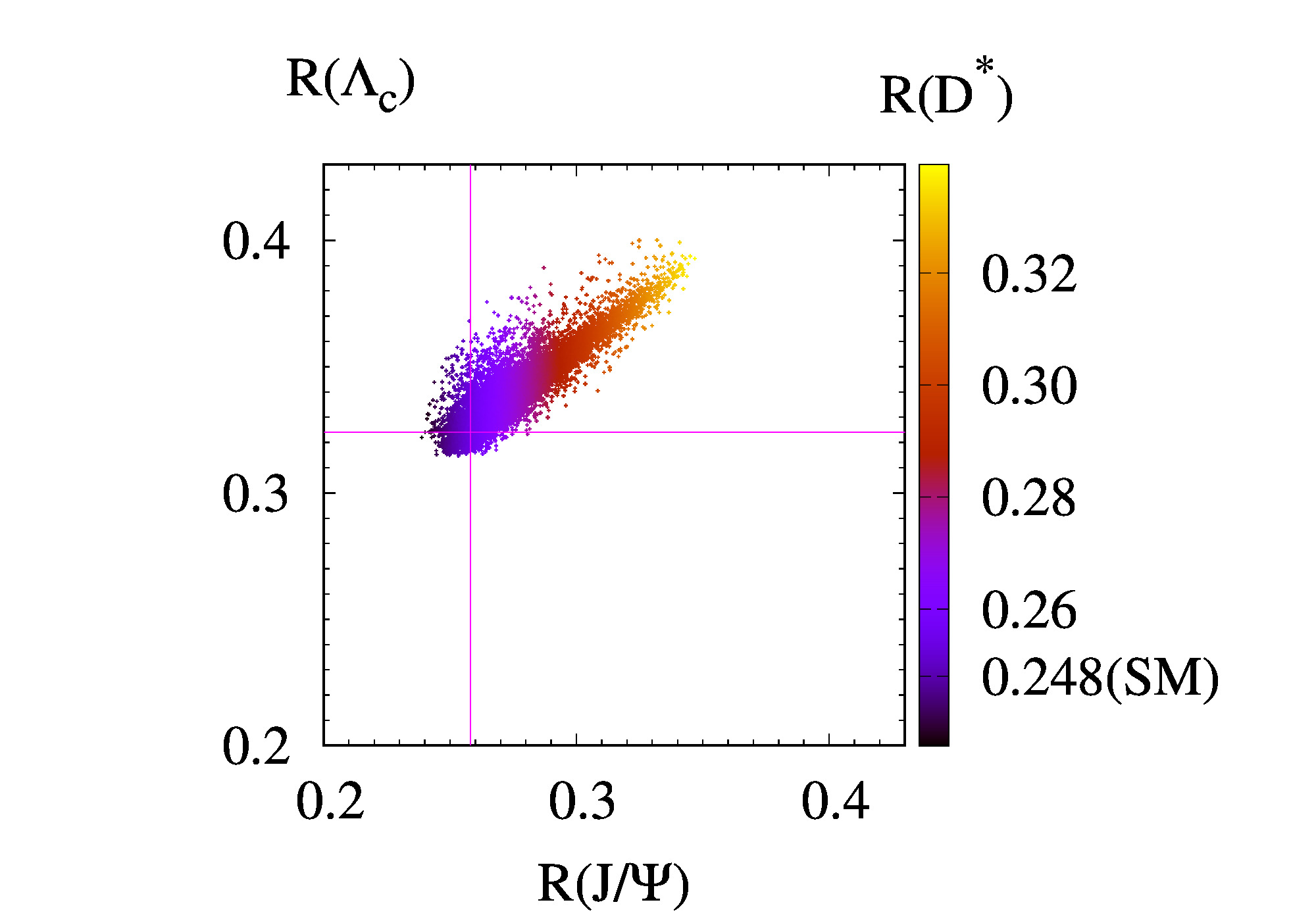} &\\
(e) &
\end{tabular}
\caption{\label{F_b2c} Allowed regions at the $2\sigma$ level for
(a) $R(D^*)$ vs. $R(D)$,
(b) $P_\tau(D^*)$ vs. $P_\tau(D)$ with respect to $F_L(D^*)$,
(c) $R(\Lambda_c)$ (green) and $R(J/\Psi)$ (blue) vs. $R(D)$, 
(d) $R(\Lambda_c)$ (green) and $R(J/\Psi)$ (blue) vs. $R(D^*)$, and
(e) $R(\Lambda_c)$ vs. $R(J/\Psi)$ with respect to $R(D^*)$.
Magenta lines are the SM values while in (c) and (d) the black line is for $R(\Lambda_c)_\SM$ and
the cyan line is for $R(J/\Psi)_\SM$.
}
\end{figure}
%
\par
Allowed regions of various observables are given in Fig.\ \ref{F_b2c}.
The SM predictions are presented by horizontal and vertical lines.
We have very similar shapes of the allowed regions for $R(D^*)$ and $R(J/\Psi)$ 
as in Fig.\ \ref{F_b2c} (a) and (c). 
Note that Eqs.\ (\ref{RDsnum}) and (\ref{RJPnum}) are quite close to each other.
In Figs.\ \ref{F_b2c} (a) and (b), the SM points of $R(D)$-$R(D^*)$ and $P_\tau(D)$-$P_\tau(D^*)$ lie
at the corners of the allowed regions. 
In Fig.\ \ref{F_b2c} (d), the SM point of $R(D^*)$-$R(J/\Psi)$ is slightly off the allowed region.
On the other hand, as in Fig.\ \ref{F_b2c} (e), the SM point of $R(J/\Psi)$-$R(\Lambda_c)$ is located
within the allowed region.
This is different feature compared to our previous work of \cite{JPL2411} where the SM point of
$R(J/\Psi)$-$R(\Lambda_c)$ lies outside.
It means that simultaneous determination of these observables would be very critical to probe NP.
Figure \ref{F_b2c} (b) reveals that $P_\tau(D)$ is definitely positive while $P_\tau(D^*)$ negative.
In \cite{JPL2411} it was shown that if $P_\tau(D)$ were negative then the shape of NP would be
restricted very strongly.
%

%
%
%
%
%
\begin{figure}
\hspace{-1cm}\includegraphics[scale=0.12]{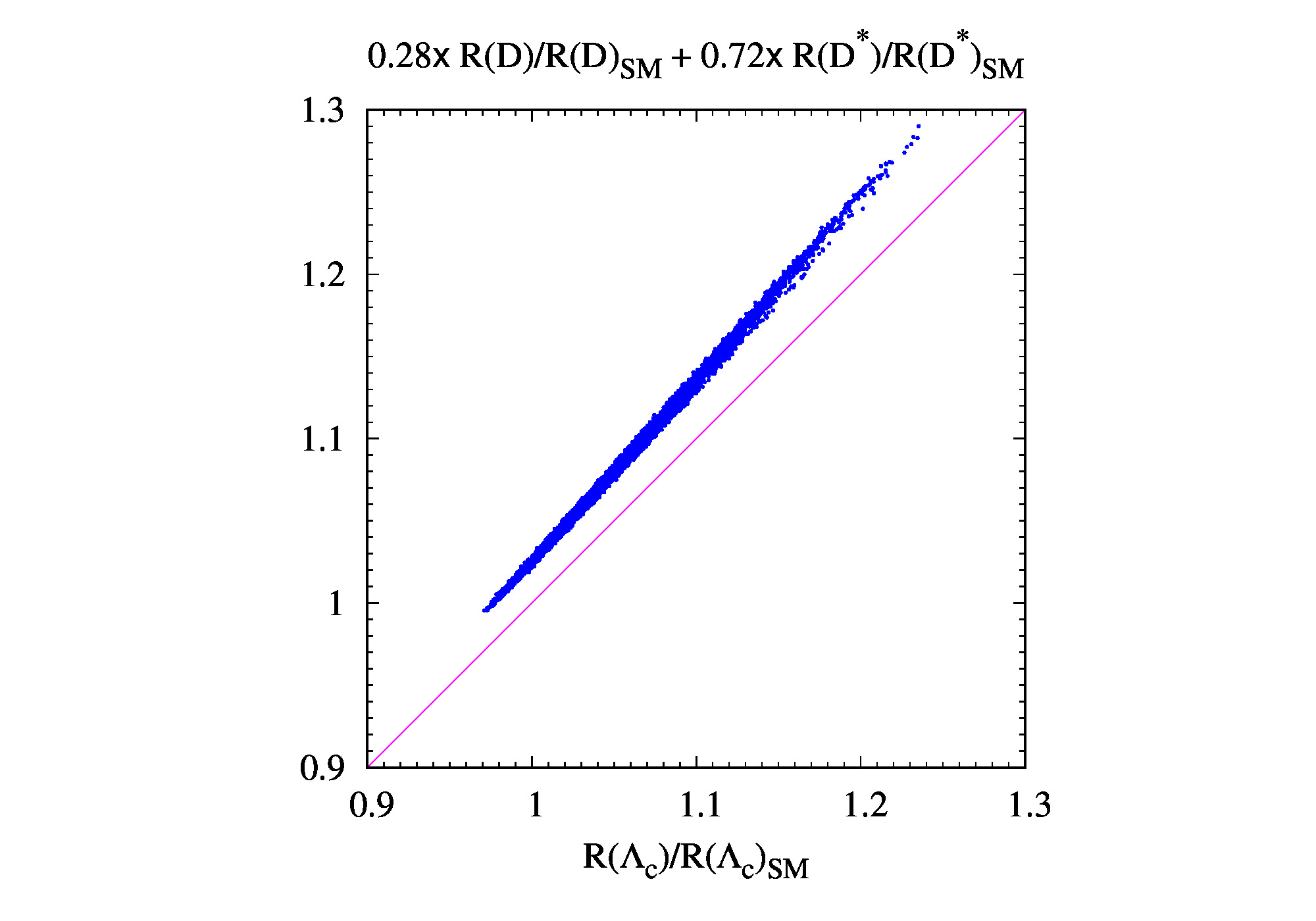}
\caption{\label{F_sumrule} Allowed regions at the $2\sigma$ level for
$0.28\frac{R(D)}{R(D)_\SM} + 0.72\frac{R(D^*)}{R(D^*)_\SM}$ vs. $\frac{R(\Lambda_c)}{R(\Lambda_c)_\SM}$
.}
\end{figure}
%
\par
In Fig.\ \ref{F_sumrule} we check the sum rule for $R(\Lambda_c)$.
As shown in the Figure one can find easily that 
\begin{equation}
\frac{R(\Lambda_c)}{R(\Lambda_c)_\SM} 
= 0.28\frac{R(D)}{R(D)_\SM} + 0.72\frac{R(D^*)}{R(D^*)_\SM} 
   + \delta_{\Lambda_c}~,
\end{equation}
with $\delta_{\Lambda_c} < 0$.
The gap $\delta_{\Lambda_c}$ tends to get larger when $R(\Lambda_c)$ increases.
A dominant term of $\delta_{\Lambda_c}$ is proportional to  $-|C_T|^2$ \cite{Iguro2405}.
In our analysis we have $|C_T|\ll 1$ as can be seen in Fig.\ \ref{F_WC} (b),
so the sum rule is not violated significantly.
And also $R(\Lambda_c)$ contains a term of $+10.4 |C_T|^2$.
As a result larger $|C_T|^2$ makes $R(\Lambda_c)$ and $-\delta_{\Lambda_c}$ larger,
which explains why the gap in Fig.\ \ref{F_sumrule} gets larger for large $R(\Lambda_c)$.
In \cite{JPL2509}, we find similar sum rule for dineutrino decay modes.
Theoretical survey is given in \cite{Kitahara2604}.
%
%
%
%
%
\begin{figure}
\begin{tabular}{cc}
\hspace{-1cm}\includegraphics[scale=0.12]{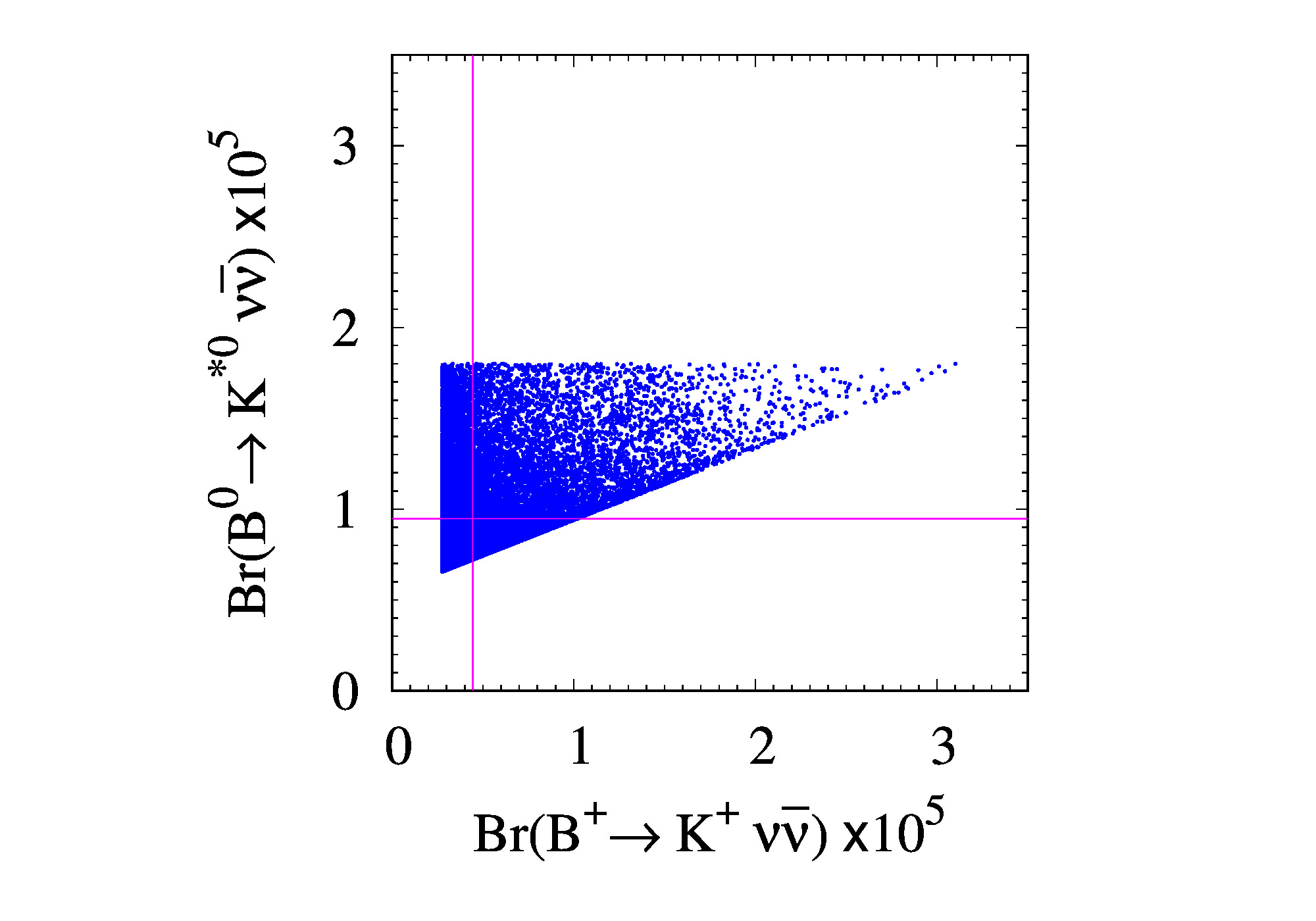}&
\hspace{-1cm}\includegraphics[scale=0.12]{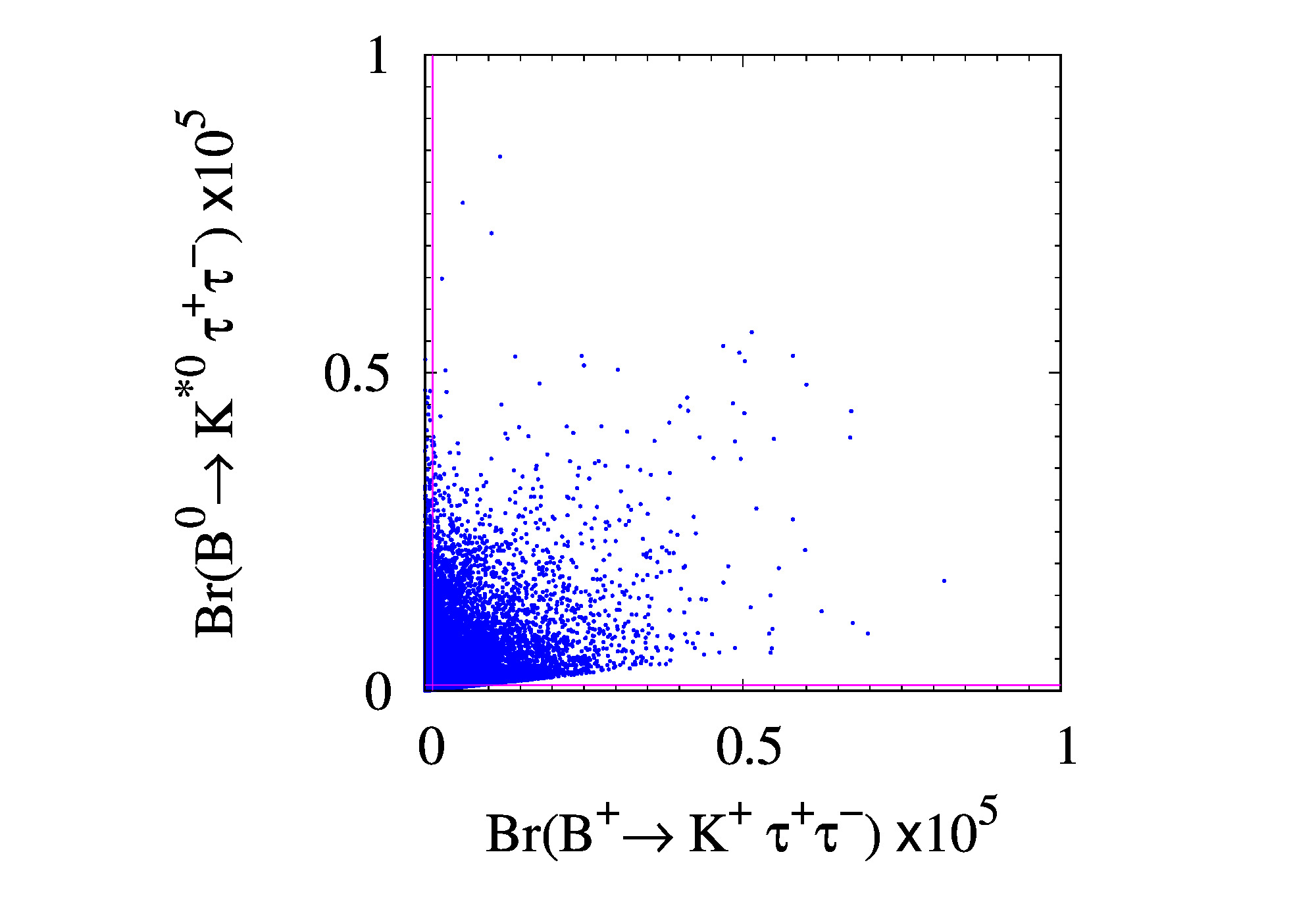}\\
(a) & (b) \\
\hspace{-1cm}\includegraphics[scale=0.12]{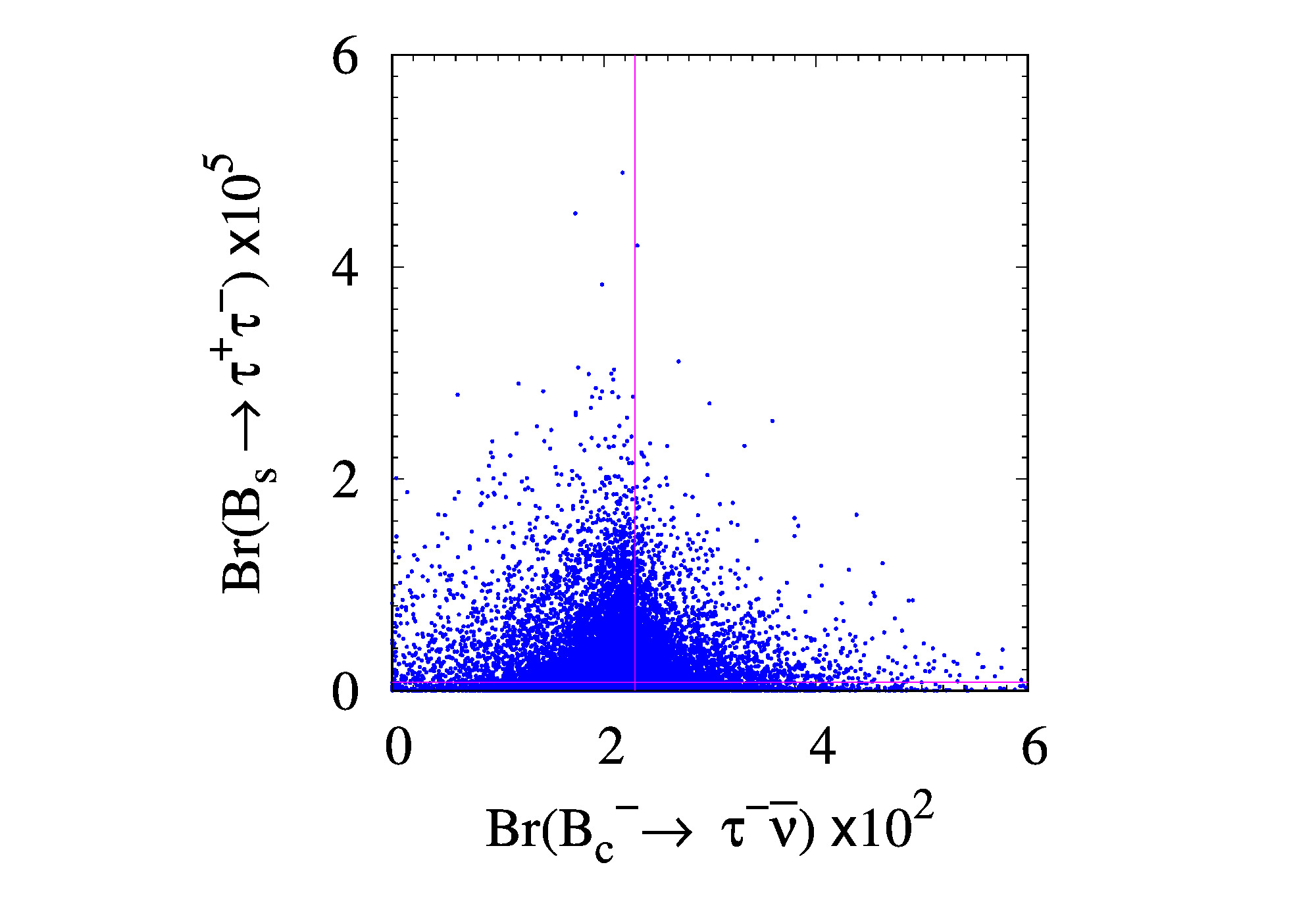}&
\hspace{-1cm}\includegraphics[scale=0.12]{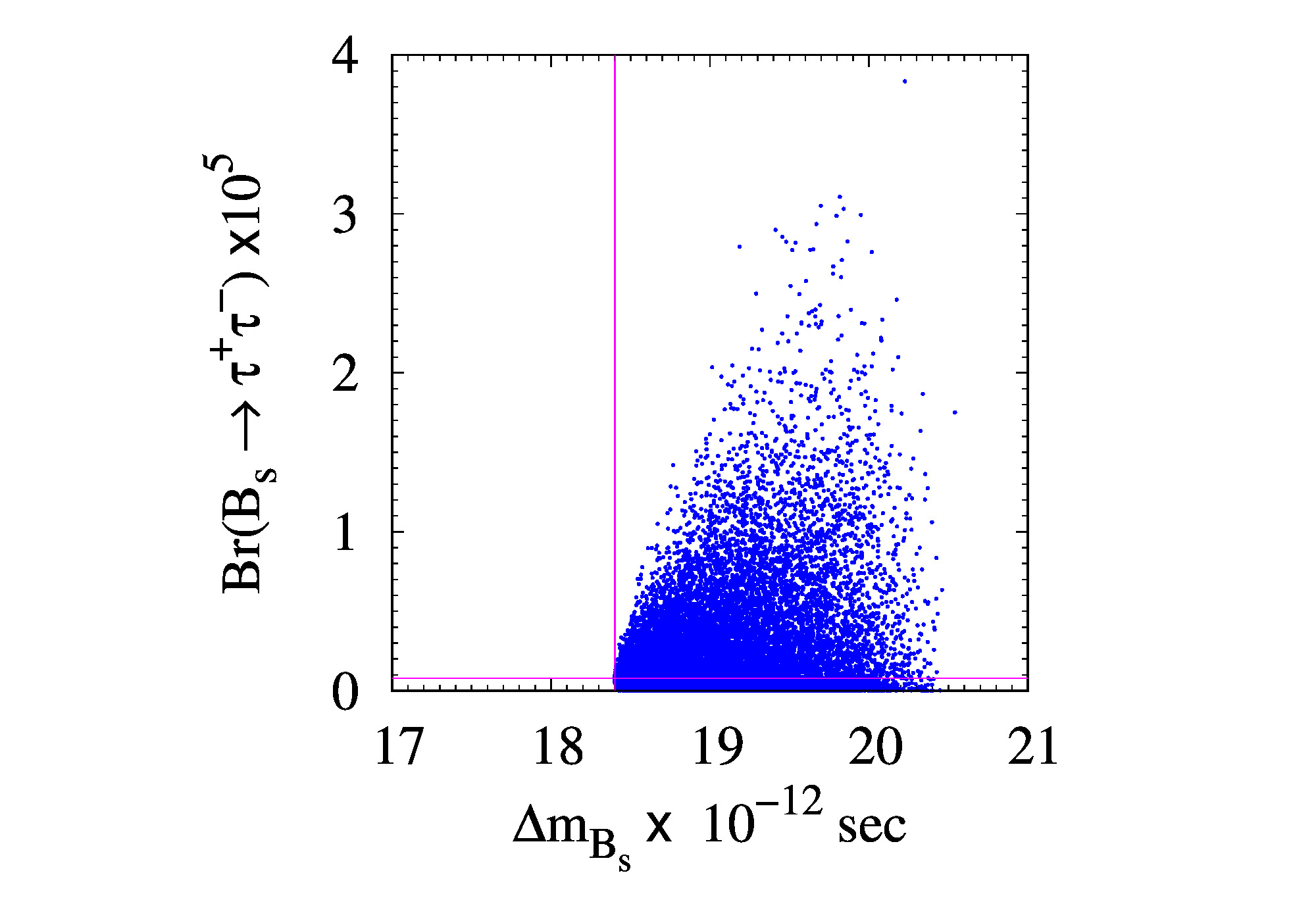}\\
(c) & (d)
\end{tabular}
\caption{\label{F_BrKKs} Allowed regions at the $2\sigma$ level for
(a) $\Br(B^0\to K^{*0}\nu\nubar)$ vs. $\Br(B^+\to K^+\nu\nubar)$,
(b) $\Br(B^0\to K^{*0}\tau^+\tau^-)$ vs. $\Br(B^+\to K^+\tau^+\tau^-)$,
(c) $\Br(B_s\to\tau^+\tau^-)$ vs. $\Br(B_c^-\to\tau^-\nubar)$, and
(d) $\Br(B_s\to\tau^+\tau^-)$ vs. $\Delta m_{B_s}$.
Magenta lines represent the SM values.
}
\end{figure}
%
\par
Figure \ref{F_BrKKs} depicts the branching ratios, 
$\Br(B\to\Ks\nu\nubar)$, $\Br(B\to\Ks\tau^+\tau^-)$, $\Br(B_c^-\to\tau^-\nubar)$, $\Br(B_s\to\tau^+\tau^-)$,
and $\Delta m_{B_s}$.
As before the SM predictions are given by magenta lines.
Note that $\Br(B^0\to K^{*0}\nu\nubar)$ and $\Br(B\to \Ks\tau^+\tau^-)$ are yet to be measured.
There are still big gaps of a few orders between experimental bounds and the SM estimations especially in
$b\to s \tau^+\tau^-$ decays.
While the allowed ranges of the branching ratios are embracing well the SM values, 
$\Delta m_{B_s}^\SM$ lies on the verge of the  allowed region.
Since $\Delta m_{B_s}^\exp < \Delta m_{B_s}^\SM$ and
individual LQs tend to constructively contribute to $\Delta m_{B_s}$, 
the mixing term of different LQs are very important to fit the data.
\par
Experimentally, our best-fit values of the LQ masses 
$M_{S_1}^\best\sim M_{\tR_2}^\best < M_{S_3}^\best \lesssim 3$ TeV 
will be challenged by the High-Luminosity LHC (HL-LHC) of $3000~ {\rm fb}^{-1}$
in channels where LQ decays into $\tau$ and $b$.
The discovery potential of the HL-LHC for scalar LQs is expected to go up to $1.2\sim 1.7$ TeV, 
while the exclusion sensitivity is $1.4\sim 1.9$ TeV \cite{ATL_22_018,CMS_22_001}.
Unfortunately $M_{S_3}^\best = 2.62$ TeV is slightly beyond the scope of the HL-LHC.
%
%
%
%
%
\section{Conclusions}
%
%
%
%
%
We explored LQ scenarios with $S_1$, $\tR_2$, and $S_3$ scalars to explain $B$ anomalies, 
especially focusing on the tauonic sector.
Each of three scalar LQs has their own roles for $B$ anomalies.
Firstly we needed $S_1$ to explain $R(\Ds)$, $P_\tau(\Ds)$ and $F_L(D^*)$ with $C_{V_L}$, $C_{S_L}$, and
$C_T$.
It is well known that $S_1$ can contribute to $b\to s \ell^+\ell^-$ decays via charm loops, 
but they are not enough to explain the experimental data.
Fortunately $S_3$ can do the work by $C_{9,10}$ Wilson coefficients.
However the newly measured branching ratio of $B^+\to K^+\nu\nubar$ decay combined with the previous
searches of $\Br(B^0\to K^{*0}\nu\nubar)$ requires non-zero $C_R^\nu$, 
while $S_1$ and $S_3$ only contribute to $C_L^\nu$.
Here comes another scalar LQ $\tR_2$ for $C_R^\nu$.
\par
A strong constraint is imposed by $\Delta m_{B_s}$ of $B_s$-$\Bbar_s$ mixing.
Measured values of $\Delta m_{B_s}$ are very close to and slightly below the SM predictions, while
each scalar LQ contributes to $\Delta m_{B_s}$ constructively with respect to the SM 
when the relevant couplings are all real.
But there is a mixing term of $S_1$ and $S_3$ which can contribute destructively to $\Delta m_{B_s}$.
In other words, $\Delta m_{B_s}$ requires both of $S_1$ and $S_3$, not one of them only.
To fit the experimental data, negative values of mixing term are preferred.
\par
Our analyses favor subdominant $C_{V_L} < 0$ and dominant $C_{S_L} > 0$ 
while the opposite signs are also allowed.
For $C_9^\LQ$, we found that LFU violating value of $\approx +1$ is favored
and the best fit $C_9^\prime$ is $-8.37$.
In the dineutrino decay modes both of Wilson coefficients $C_{L,\LQ}^\nu$ and $C_R^\nu$
favor negatively large values, which is consistent with our previous works.
Experimental data for $b\to s\tau^+\tau^-$ are still lack and the upper bounds are orders of magnitude 
higher than the SM predictions.
Future experiments could remove the ambiguities and probe the scalar LQ masses.
%
%
%
%
%
%
\begin{acknowledgments}
This paper was supported by Konkuk University in 2024.
\end{acknowledgments}
%
%
%
%
%
%
%
%
%
%
%

\end{document}